\documentclass[runningheads]{llncs}
\usepackage[T1]{fontenc}
\usepackage{graphicx}
\usepackage{booktabs} % For formal tables
\usepackage[acronym, nowarn]{glossaries}

\makeglossaries

\usepackage{makecell}
\usepackage{multirow}

\usepackage{diagbox}
\usepackage[font=footnotesize,labelfont=bf]{caption}
\usepackage{xcolor, colortbl}
\usepackage{xspace}

\usepackage{todonotes}

\usepackage{tikz}

\usepackage{halloweenmath} % -> cloud
\usepackage{hyperref}
\usepackage{cleveref}
\usepackage{latexsym}
\usepackage{transparent}
\usepackage[final]{changes} % for final changes
\usepackage{comment}
\usepackage{adjustbox}

\usepackage{listings}
\lstnewenvironment{codenv}{\lstset{language=SQL, 
basicstyle=\footnotesize}}{}
\usepackage{xcolor}

\newcommand{\system}{\textsc{\textsc{PDSP-Bench}}\xspace} 

\usepackage[]{todonotes} % disable as flag
\newcommand{\first}{\emph{(i) }}
\newcommand{\ii}{\emph{(ii) }}
\newcommand{\iii}{\emph{(iii) }}

\newcommand{\ie}{i.e.,\xspace}
\newacronym{dsp}{DSP}{Distributed Stream Processing}
\newacronym{dcep}{DCEPS}{Distributed Complex Event Processing Systems}
\newacronym{sp}{SP}{Stream Processing}
\newacronym{sps}{SPS}{Stream Processing System}
\newacronym{dag}{DAG}{Directed Acyclic Graph}
\newacronym{gnn}{GNN}{Graph Neural Network}
\newacronym{mlps}{MLP}{Multi-Layer Perceptron}
\newacronym{qos}{QoS}{Quality of Service}
\newacronym{iot}{IoT}{Internet of Things}
\newacronym{nic}{NIC}{Network Interface Cards}
\newacronym{ml}{ML}{Machine Learning}
\newacronym{pqp}{PQP}{parallel query plans}
\newacronym{inp}{INP}{In-Network Processing}
\newacronym{sut}{SUT}{System under Test}
\newacronym{wui}{WUI}{Web User Interface}
\newacronym{udo}{UDOs}{user-defined operators}
\newacronym{pdsp}{PDSP}{parallel and distributed stream processing}
\usepackage{color}
\usepackage{listings}
\usepackage{graphicx} % Required for including images
\usepackage{subcaption} % Required for creating sub-figures
\newboolean{komversion}
\setboolean{komversion}{true} % set true to print KOM-headers and footers
\ifthenelse{\boolean{komversion}}{
    \usepackage{eso-pic}% http://ctan.org/pkg/eso-pic
    \AddToShipoutPictureBG{% Add picture to background of every page
        \AtPageUpperLeft{%
            \raisebox{-3\baselineskip}{
\makebox[\paperwidth]{\colorbox{yellow!40}{\begin{minipage}{0.85\paperwidth}\centering\small
                            {                            Pratyush Agnihotri, Boris Koldehofe, Roman Heinrich, Carsten Binnig, Manisha Luthra. PDSP-Bench: A Benchmarking System for Parallel and Distributed Stream Processing.  To Appear in Proceedings of the Performance Evaluation and Benchmarking - 16th {TPC} Technology Conference,
                  {TPCTC} 2024, Springer, 22 pages}
                        \end{minipage}}}}%
                    }
                    \AtPageLowerLeft{%
                        \raisebox{3\baselineskip}{
\makebox[\paperwidth]{\fbox{\begin{minipage}{0.85\paperwidth}\scriptsize
                                        {The documents distributed by this server have been provided by the contributing authors as a means to ensure timely dissemination of scholarly and technical work on a non-commercial basis. Copyright and all rights therein are maintained by the authors or by other copyright holders, not withstanding that they have offered their works here electronically. It is understood that all persons copying this information will adhere to the terms and constraints invoked by each author's copyright. These works may not be reposted without the explicit permission of the copyright holder.}
                                    \end{minipage}}}}%
                    }
    }
}{} 
\begin{document}
\title{\system: A Benchmarking System for\\Parallel and Distributed Stream Processing}
\author{Pratyush Agnihotri\inst{1} \and
Boris Koldehofe\inst{2} \and
Roman Heinrich \inst{1,3} \and \\
Carsten Binnig \inst{1,3} \and
Manisha Luthra \inst{1,3}
}
\institute{
Technische Universität Darmstadt, Germany \and
%\email{\{firstname,lastname\}@kom.tu-darmstadt.de} \and
Technische Universität Ilmenau, Germany \and
%\email{\{firstname,lastname\}@tu-ilmenau.de} \and
DFKI Darmstadt, Germany \\
%\email{firstname,lastname\}@dfki.de}
}

\maketitle              % typeset the header of the contribution
%
%\begin{abstract}
%The abstract should briefly summarize the contents of the paper in

%\keywords{First keyword  \and Second keyword \and Another keyword.}
%\end{abstract}
\vspace{-3.0ex}
\begin{abstract}
  The paper introduces~\system, a novel benchmarking system designed for a systematic understanding of performance of parallel stream processing in a distributed environment. 
Such an understanding is essential for determining how Stream Processing Systems (SPS) use operator parallelism and the available resources to process massive workloads of modern applications. % concurrently and efficiently. 
%and the effective utilization of heterogeneous resources in processing diverse and high-volume workloads. 
Existing benchmarking systems focus on analyzing SPS using queries with sequential operator pipelines within a homogeneous centralized environment. 
Quite differently,~\system emphasizes the aspects of parallel stream processing in a distributed heterogeneous environment and simultaneously allows the integration of machine learning models for SPS workloads. %, which becomes essential for SPS given the modern demands of today's applications.
%yet challenging to measure on diverse workloads and hardware configurations. 
%Additionally, \system offers interfaces for training data generation and inferences using ML approaches, catering the growing trend of automatically tuning parallelism and resources.  
%introduces a cost analysis benchmark for Machine Learning (ML) models within streaming contexts. 
In our results, we benchmark a well-known SPS, Apache Flink, using parallel query structures derived from real-world applications and synthetic queries to show the capabilities of~\system towards parallel stream processing. 
Moreover, we compare different learned cost models using generated SPS workloads on \system by showcasing their evaluations on model and training efficiency.  
%We show anecdotes from our experiments using \system that highlight interesting trends given different query workloads and hardware complexity on the performance of a stream processing system. 
We present key observations from our experiments using \system that highlight interesting trends given different query workloads, such as \textit{non-linearity} and \textit{paradoxical} effects of parallelism on the performance. % of a \acrshort{sps}.  %stream processing system.
%complexities and heterogeneous hardware ‚
%to assess SPS %performance and scalability on heterogeneous hardware configuration in terms of Quality of Service and resource utilization, 
%which are critical for parallel stream processing. 
\end{abstract}

% Introduction

\glsresetall
\vspace{-6ex}
% ============================================
\section{Introduction}
\label{sec:intro}
% ============================================

\begin{table*}[]
\tiny
\begin{tabular}{ccccccccccc}
\hline
                                          &                                                                                                      &                                                                                                           & \multicolumn{1}{l}{}                                                                                              &                                                                                     &                                                                                      & \multicolumn{2}{c}{\textbf{Application Suite}}                                &                                                                                                                                   \\ \cline{7-8} \cline{10-11} 
\multirow{-2}{*}{\textbf{\begin{tabular}[c]{@{}c@{}}Benchmark\\System\end{tabular}}}      & \multirow{-2}{*}{\textbf{\begin{tabular}[c]{@{}c@{}}C1: P/S\end{tabular}}} & \multirow{-2}{*}{\textbf{\begin{tabular}[c]{@{}c@{}}C2: He/Ho\end{tabular}}} & \multicolumn{1}{l}{\multirow{-2}{*}{\textbf{\begin{tabular}[c]{@{}l@{}}D/C\end{tabular}}}} & \multirow{-2}{*}{\textbf{Infrastructure}}                                           & \multirow{-2}{*}{\textbf{\begin{tabular}[c]{@{}c@{}}C3: Learned \\ SPS Models\end{tabular}}} & \multicolumn{1}{l}{\textbf{Real-world}} & \textbf{Synthetic}                  & \multirow{-2}{*}{\textbf{Scalability}} \\ \hline
Linear Road~\cite{arasu2004linear}                              & S                                                                                                    & Ho                                                                                                        & C                                                                                                                 & \begin{tabular}[c]{@{}c@{}}Single\\machine\end{tabular}                                                                      & No                                                                                   & 1                                       & -                                   & No                                                                                     \\ \hline
YSB~\cite{yahooChintapalli2016benchmarking}                                       & S                                                                                                    & Ho                                                                                                        & C                                                                                                                 & \begin{tabular}[c]{@{}c@{}}Single\\machine\end{tabular}                                                                      & No                                                                                   & 1                                       & -                                   & No                                                                                                                         \\ \hline
StreamBench~\cite{lu2014streamBench}                               & S                                                                                                    & Ho                                                                                                        & D                                                                                                                 & VMs                                                                                 & No                                                                                   & -                                       & 7                                   & Partially                                                                                                          \\ \hline
RIoTBench~\cite{shukla2017riotbench}                                 & S                                                                                                    & Ho                                                                                                        & D                                                                                                                 & VMs                                                                                 & No                                                                                   & 4                                       & -                                   & No                                                                                                                         \\ \hline
OSPBench~\cite{van2020evaluationOSPBench}                                  & S                                                                                                    & Ho                                                                                                        & D                                                                                                                 & \begin{tabular}[c]{@{}c@{}}Cloud\\ AWS EC2\end{tabular}                                                                       & No                                                                                   & -                                       & 1                                   & No                                                                                                                       \\ \hline
HiBench~\cite{huang2010hibench}                                   & S                                                                                                    & Ho                                                                                                        & D                                                                                                                 & \begin{tabular}[c]{@{}c@{}}Local\\ Cluster\end{tabular}                                                                       & No                                                                                   & -                                       & 4                                   & No                                                                                                                        \\ \hline
BigDataBench~\cite{wang2014bigdatabench}                              & S                                                                                                    & Ho                                                                                                        & D                                                                                                                 & \begin{tabular}[c]{@{}c@{}}Local\\ Cluster\end{tabular}                                                                       & No                                                                                   & -                                       & 1                                   & Partially                                                                                                                  \\ \hline
ESPBench~\cite{hesse2021espbench}                                  & S                                                                                                    & Ho                                                                                                        & D                                                                                                                 & VMs                                                                                 & No                                                                                   & 5                                       & -                                   & No                                                                                                                 \\ \hline
SPBench~\cite{garcia2023spbench}                                   & P                                                                                                    & Ho                                                                                                        & C                                                                                                                 & VMs                                                                                 & No                                                                                   & 4                                       & -                                   & Partially                                                                                                         \\ \hline
DSPBench~\cite{bordin2020dspbench}                                  & P                                                                                                    & Ho                                                                                                        & D                                                                                                                 & \begin{tabular}[c]{@{}c@{}}Azure\\ Cloud Cluster\end{tabular}                                                                 & No                                                                                   & 13                                      & 2                                   & Partially                                                                                                          \\ \hline
\cellcolor[HTML]{EFEFEF}\textbf{\system} & \cellcolor[HTML]{EFEFEF}\textbf{S/P}                                                                   & \cellcolor[HTML]{EFEFEF}\textbf{He/Ho}                                                                    & \cellcolor[HTML]{EFEFEF}\textbf{C/D}                                                                                & \cellcolor[HTML]{EFEFEF} \textbf{\begin{tabular}[c]{@{}c@{}}CloudLab,\\Geni Cluster,\\ On-premise\end{tabular}} & \cellcolor[HTML]{EFEFEF}\textbf{Yes}                                                 & \cellcolor[HTML]{EFEFEF}\textbf{14}     & \cellcolor[HTML]{EFEFEF}\textbf{9} & \cellcolor[HTML]{EFEFEF}\textbf{Fully}                                                   \\ \hline
\end{tabular}
\vspace{0.5ex}
\caption{Comparison of the existing benchmarking system for \acrshort{sp} with \system emphasizing the research challenges. Our work can effectively benchmark both parallel data flow graphs and heterogeneous hardware as well as can be used as a benchmarking system for training ML models on \acrshort{sp} workloads. Abbreviations used are S: Sequential plans, P: Parallel plans, He: Heterogeneous hardware, Ho: Homogeneous hardware, D: Distributed cluster, and C: Centralized or single machine.}
\vspace{-3ex}
\label{tab:relatedWork}
\end{table*}
%\todo{fix YSB reference in table and check other references}
\textbf{Benchmarking parallel dataflows is important.} Recent advancements in \acrfull{sp} have introduced a variety of systems, such as Apache Flink and Storm for analyzing data streams in real-time. 
These systems are essential for many applications that handle immense data volumes. 
For instance, Netflix uses Flink to process over $1.3TB$ of data in its daily tasks, necessitating the use of many parallel operator instances to keep up with the high arrival rates and processing of data tuples~\cite{netflixFlink}. 
For this, \acrshort{sp} systems offer a \emph{data flow} abstraction to specify operator parallelism in the query and provide data partitioning strategies to manage data stream partitions. 
While such \textit{parallel data flows} have become an intrinsic part of every \acrshort{sp} system,  
there is, however, very limited understanding of the performance of \acrshort{sps} under massively parallel dataflows. 

\textbf{Key challenges of existing work.} Most of the existing benchmarking systems for \acrshort{sp} are tailored towards the understanding of sequential dataflows \cite{arasu2004linear,yahooChintapalli2016benchmarking,wang2014bigdatabench,Asterios2018benchmark}. 
Those benchmarking parallel dataflows~\cite{van2020evaluationOSPBench,garcia2023spbench,bordin2020dspbench,zeuch2019analyzing} are restricted to a homogeneous environment for resource placement and offer limited capabilities in terms of scaling workloads, e.g., event rate and query parameters like window length. 
We believe a thorough analysis of parallel data flow graph placement on heterogeneous resources will reveal interesting insights into the behavior of distinct operators on various hardware resources and vice versa. 
Another unique aspect of our work is the ability to scale workload generation -- both data streams and queries -- by offering a benchmarking platform, which in fact can also be used for machine learning of \acrshort{sps} workloads, becoming increasingly important nowadays~\cite{heinrich2022debs,zapridou2022dalton}. 
In summary, we identify three primary challenges for \system by analyzing key existing benchmarking systems for \acrshort{sp} workloads presented in \Cref{tab:relatedWork}.% which we summarize and exemplify as follows. «

\textbf{C1: Lack of expressiveness.} Most existing benchmarks~\cite{yahooChintapalli2016benchmarking,arasu2004linear,wang2014bigdatabench,huang2010hibench} often overlook the importance of benchmarking parallel dataflow applications, thus focusing only on sequential dataflows with a limited set of operators. 
For instance, StreamBench~\cite{lu2014streamBench} overlooks essential operators, such as window functions, crucial for concurrent partitioning and efficient resource utilization.

\textbf{C2: Shift to heterogeneity.} The shift towards 
%distributed and 
heterogeneous hardware requirement for benchmarking requires complex resource management, i.e., the underlying system must manage parallel resource mapping on varied hardware architectures, network links, and storage. 
Although benchmarking systems exist that assess parallel dataflows, like DSPBench~\cite{bordin2020dspbench} and SPBench~\cite{garcia2023spbench}, the benchmarks are restricted to homogeneous hardware 
reducing their relevance as real-world workloads often require heterogeneous environments~\cite{zeuch2019analyzing}. For instance, Netflix runs on $1400+$ nodes on $50+$ distinct clusters with varied CPU cores~\cite{netflixFlink} to deal with their demands of 
massively parallel dataflow applications. 

\textbf{C3: Integrating learned \acrshort{sps} models.} Most importantly, the rapid advancement in \acrshort{sp} mechanisms using machine learning (ML), necessitates a scalable and resource-friendly benchmarking system, ensuring its long-term relevance and utility in assessing future \acrshort{sps} with learned components. 
Recently, ML has been successfully applied for cost-based optimizations in SPS to support heterogeneous placements~\cite{heinrich2022debs,heinrich2024costream} and deciding parallelism strategies~\cite{zapridou2022dalton,PrAg_ZeroTune_ICDE_2024} and showed promising performance. This increasing surge of development in learned SPS models calls for a benchmarking platform that allows fair comparison between them by integrating the models and generating consistent training data for them. 
%internal 
%ML models become increasingly integrated into 
%\acrshort{sps} mechanisms like resource management~\cite{heinrich2022debs} and parallelism tuning~\cite{zapridou2022dalton}. 
However, existing work 
do not provide a means to integrate ML models such that they can be compared in a ``fair'' way with consistent metrics (cf.~\Cref{tab:relatedWork}, Learned SPS models).%ML integration). 
%even fail to provide the basic functions required to collect training data for ML models, e.g., generating data streams with different event rates. % for incoming data streams. 

%Moreover, the applicability of such ML models not only requires synthetic data generation but also a means to enumerate real-world streaming applications. 
%In fact, none of the existing benchmarks (cf.~\Cref{tab:relatedWork}, column ML integration) provide these basic functions, severely slowing down the advancement for ML in \acrshort{sps}.
%however, existing benchmarks often fail to provide basic functions like scaling workloads and ability to test heterogeneous hardware slowing down the advancement for ML in \acrshort{sps}.

\textbf{Our proposal:} 
We propose \system\footnote{Source code: \url{https://github.com/pratyushagnihotri/PDSPBench}}, a novel benchmarking system specifically designed to tackle the three primary challenges faced by existing benchmarks: lack of expressiveness in benchmarking parallel dataflows, the necessity for heterogeneous hardware support, and the integration of learned SPS models. 
Unlike existing benchmarks, \system enables the creation and evaluation of parallel query structures (PQP) across a diverse range of operators and input data streams, which we divide into synthetic and real-world workloads, thus offering an expressive and scalable solution. 
We also provide mechanisms to configure and manage heterogeneous hardware resources, by integrating resources from testbeds like CloudLab with different configurations, which are essential for accurately reflecting real-world deployment scenarios.
Furthermore, \system facilitates the integration of learned SPS models, allowing for systematic training and evaluation of these models on diverse streaming workloads. This integration is increasingly important given the surge of use of ML for optimizing SPS performance~\cite{PrAg_ZeroTune_ICDE_2024,heinrich2024costream,zapridou2022dalton}.  
The system's ability to generate large corpora of streaming datasets ensures that the ML models are trained on data representative of actual streaming workloads. (cf. \Cref{sec:pspBench_PQP_IDG})

The evaluation of \system involves extensive experiments that highlight its capabilities in benchmarking parallel stream processing. 
By using Apache Flink as the System Under Test (SUT), the evaluation demonstrates the impact of varying parallel query complexities, hardware configurations, and workload parameters on system performance. 
The results show the importance of considering both parallelism and heterogeneity to achieve optimal performance in real-time data processing applications. 
Moreover, we also integrate and train various learned cost models for streaming queries and showcase their performance in terms of model and training efficiency. (cf. \Cref{sec:results})%\todo{need to mention ML evaluation}

\vspace{-3ex}
\section{\system: System Overview}\label{sec:systemOverview}
\begin{figure}
    \centering
        \includegraphics[width=0.7\linewidth]{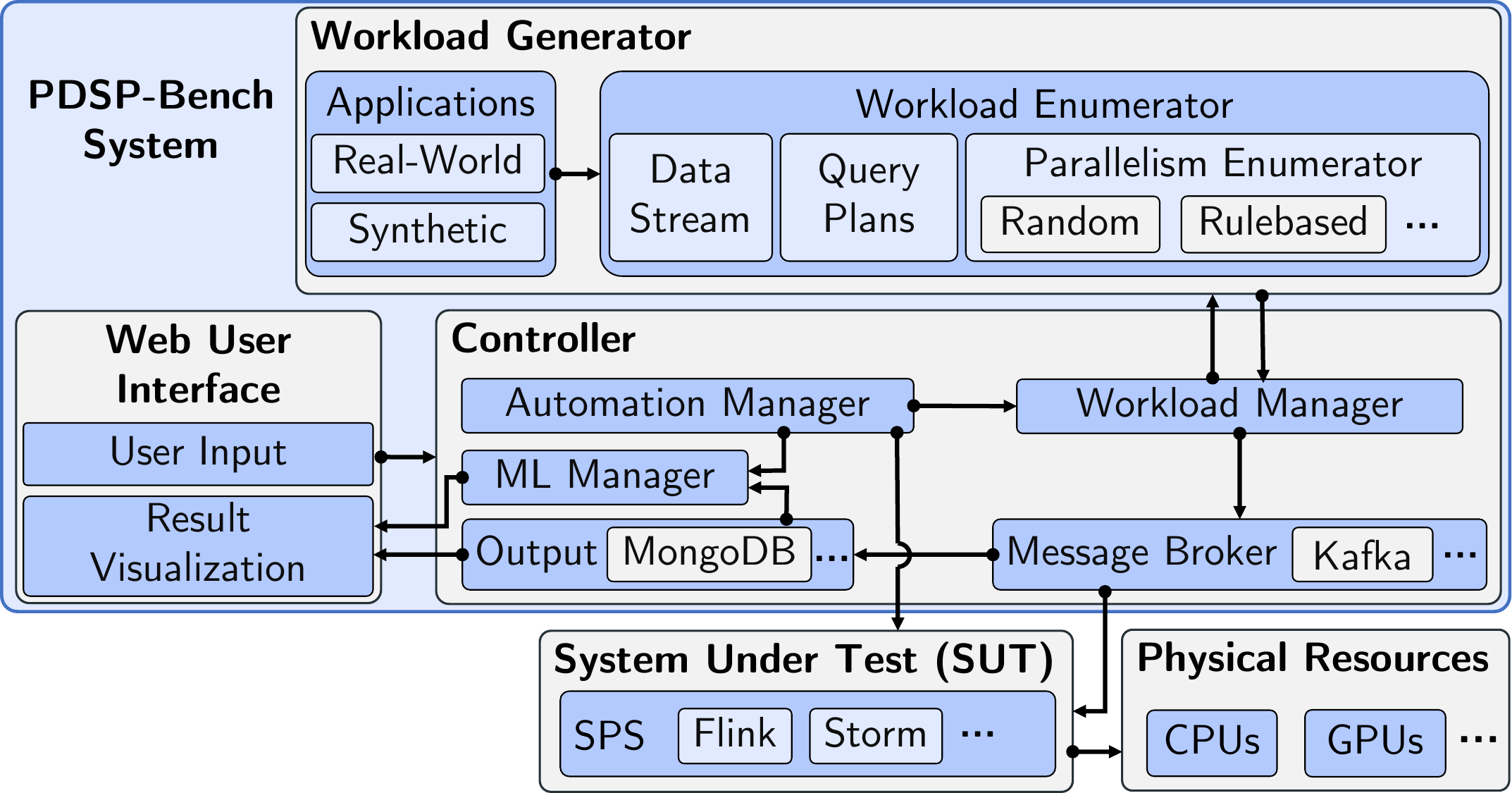}
        \label{fig:sub1}
        \vspace{-2ex}
    \caption{\system system overview}
    \label{fig:pspbench_overview}
\end{figure}
The main goal of \system is to enable benchmarking of~\acrfull{pdsp} systems considering heterogeneous environments for query deployment. 
As such, \system aims to enable the creation of large corpora of streaming datasets across three dimensions: \textit{query}, \textit{data} and \textit{resource} diversity. 
Such large corpora of datasets can be used in training ML models for learning optimizations of \acrshort{sps} such as cost of executing streaming queries and their placement on heterogeneous hardware. 
We demonstrate this by training and evaluating learned cost models using the dataset generated by \system. 
%In the following, we provide an overview of~\system designed for benchmarking~\acrshort{sps} for~\acrfull{pdsp} considering heterogeneous environments. 

While \system supports both sequential and parallel query plans (PQPs)\footnote{By PQP, we mean a given query structure with parallelism degrees that can generate multiple queries of this type of structure, e.g., linear PQP will generate a plan with parallel instances of filter operators with random filter literals.}, we mainly focus on \acrshort{pqp} to show our novel contributions to tackle challenges. 
\system has three main components: (1) {workload generator}, (2) {controller} and (3) {web user interface} (cf.~\Cref{fig:pspbench_overview}). 
\acrfull{sut} represents the underlying~\acrshort{sps} like Apache Flink or Storm that are being evaluated by \system. 
We present an overview of our solution (\textbf{S\#}) towards the goal and show how we address the aforementioned challenges (\textbf{C\#}) of existing work using \system components as follows.
%Additionally, we explain how we overcome the aforementioned challenges (C1-C3).

\textbf{C1: Lack of expressiveness.} 
\textbf{S1:} To specify \acrshort{pqp} with a wide range of operators and input data streams, \system provides a core component known as {workload generator} as seen in \Cref{fig:pspbench_overview}. The task of this component is to enumerate different factors of workload -- both data and query -- e.g., \textit{parallelism degrees} to generate meaningful \acrshort{pqp} to be executed on \acrshort{sut}, e.g., on Flink, hence enabling \textit{query} and \textit{data} diversity. 
These inputs on the enumeration can be given by the user via the {web user interface} that is managed by the {controller} as discussed later, but can be also configured directly into \system. 
A \textit{key} issue we solve thereby is to generate \acrshort{pqp} that are both \textit{valid} and \textit{representative} of current streaming applications.  
Thus \acrshort{pqp} must represent both standard streaming and user-defined operators that we selected from open-source data stream processing datasets like DEBS Grand Challenges. 
We believe a combination of synthetic and real-world workloads is necessary to be able to properly assess \acrshort{sut}'s performance and generate datasets that are representative for ML in streaming platforms. 
We discuss this component in~\Cref{sec:pspBench_PQP_IDG}.

%\vspace{-0.5ex}
\textbf{C2: Shift to heterogeneity.}
\textbf{S2:} We provide interfaces to the users to configure hardware resources that are used in turn to execute the generated \acrshort{pqp} workload created by the former component. The {controller and web user interface (WUI)} components alleviate this complexity of configuring different hardware and hence enabling \textit{resource} diversity for query execution and their deployment by automating it. 
We support the evaluation of heterogeneous CPU architectures such as Intel, AMD but also distinct network, memory and storage parameters by integrating CloudLab cluster, but other cloud providers can also be integrated easily. 
Thus, the complex mechanism of creation of machines and query deployment using hefty resource providers like Kubernetes and Yarn in \acrshort{sut} is hidden using these components.

\textbf{C3: Integrating learned \acrshort{sps} models.} 
\textbf{S3:} The entire benchmarking system design holistically guides the users to specify \acrshort{pqp} and its properties as well as their execution on different hardware resources that in turn can used to generate data to train and evaluate ML models. For instance, such \acrshort{pqp} execution data can be used as features together with the performance metrics as labels, such as end-to-end latency, on a given \acrshort{sut} to train a \textit{cost model} that predicts those metrics. Moreover, {controller} component allows integration of different ML models to support training on different sizes of \acrshort{sp} workloads. To evaluate models, we report metrics such as accuracy (q-error) and training overhead (queries and time) as well as investigate trade-offs between them. 

As a solution, we present the \system workflow that shows how to use \system (cf.~\Cref{fig:pspbench_overview}) to generate streaming workloads that can be used to train ML models and in turn also be used to infer on a \acrshort{pqp} from \system using the trained model. 
All the user inputs are collected using the {\acrshort{wui}} that are forwarded to the {controller} to orchestrate the benchmarking process. 
It allows users to select from existing applications in the suite (real-world or synthetic), but also provides a means to create novel applications in the form of \acrshort{pqp}. 
Moreover, we provide other input parameters like parallelism enumeration strategies, workload and execution parameters such as event rate and query execution time (to limit the query as they are long-running) explained in the next section. 
We also allow to store the generated workload in a database, e.g., MongoDB that can be used for training ML models. 
Thus, {ML Manager} in the {controller} uses the ``same'' training data to train available ML models, e.g., a cost model can be trained to predict the costs of a \acrshort{pqp}. 
This integrated approach allows ``fair'' comparison between ML models using our reported metrics such as training overhead. 
Thus, the reporting of benchmarks must also support training- and inference-related metrics and not just performance metrics. 
%and hence facilitates {ML} advancement in streaming.
During the execution of \acrshort{pqp} on the selected \acrshort{sut}, the performance as well as the training metrics can be visualized in real-time on {\acrshort{wui}}. 
%simplifies the benchmarking process for users and facilitates {ML} advancement in streaming.

%Having provided an system overview as seen in \Cref{fig:pspbench_overview}, 
We will focus on the {workload generator} component that is core to the \system system in the following.
%for the creation of streaming datasets.  %   optimize performance and improve benchmarking outcomes.

\vspace{-3ex}
\section{Workload Generator}\label{sec:pspBench_PQP_IDG}
%To tackle~\textit{C1-C3}, 
\vspace{-2.0ex}
An important research question we answer in this work is \textit{``how to systematically generate workloads (data and query) for a comprehensive suite of dataflows to benchmark parallel and distributed streaming capabilities of~\acrshort{sut}''}. 
The {workload generator} component plays a pivotal role in this question, as it generates data streams and~\acrfull{pqp} derived using an enumerator (\Cref{subsec:workload_enum}) for our integrated synthetic and real-world applications aiding to benchmark any given~\acrshort{sut} (\Cref{subsec:applications})\footnote{Selected data: \url{https://github.com/pratyushagnihotri/pdsp-bench_experiment_data}}. 
While we generate workload by varying parameters related to data, query and resources given in \Cref{tab:pspbench_evaluation_parameters}, e.g., event rates of upto $4$ million events per second and parallelism degrees upto $128$, they are in practice limited by the amount of resources which are available (e.g., the CloudLab cluster nodes m510, c630 and c6525\_25g). 
Thus the scale of workloads that \system can be much higher given the availability of the high amount of resources. %, but we are restricted in this amount by the open-source research testbed CloudLab. 
In the following, we focus on how we diversify across these parameters.  

%By allowing these broad ranges of testing scenarios, the generator ensures that the benchmarking process provides a thorough evaluation of the~\acrshort{sut}'s performance across numerous parallel processing scenarios dictated by the applications . 
%Simultaneously, it provides an ability to \textit{enumerate} over different data stream and query ranges for both comprehensive and representative training data collection for ML models . 
\vspace{-3.5ex}
\subsection{Workload Enumerator} \label{subsec:workload_enum}
\vspace{-2.0ex}
This component enables \textit{data} and \textit{query} diversity by orchestrating the generation and a variety of data streams and~\acrshort{pqp} as described 
%As detailed in the previous section \system allows data enumeration for both real-world and synthetic applications. 
%We describe the strategies we use to enumerate data streams and query plans 
in the following. 
%It leverages different \textit{parallelism enumeration strategies} to investigate a range of \acrshort{pdsp} scenarios, ensuring workloads are accurately matched with their respective applications on the~\acrshort{sut}. 
%Additionally, it generates comprehensive \textit{workload profiles}, combining~\acrshort{pqp} with performance data, which are essential for integrating machine learning (ML) into~\acrshort{sps}, addressing \textit{C3}. 
%These profiles provide a foundation for the \textit{ML Manager} to emulate real-world operational challenges, such as varying data volumes and workload dynamics. 
%This capability is crucial for training ML models to devise strategies for effective resource management and parallelism optimization, enhancing the SPS's adaptability and efficiency. 

\textit{\textbf{Data stream:}} 
%Real-world applications often provide unstructured data that is noisy  that has to be transformed 
%may not cover all possible scenarios or edge cases that~\acrshort{sps} might encounter, making it hard to systematically assess~\acrshort{sut} performance across varying conditions. 
For synthetic applications, a common strategy to generate synthetic data is to \textit{randomly} select from a given valid data range to avoid exhaustive enumeration of the given parameters, which is extremely time consuming and practically impossible to do within a reasonable timeframe. 
In fact, \textit{domain randomization} is a common technique used for synthetic data generation to train ML models like deep neural networks such that it learns from the features of interest. 
The rationale behind this method is to have variability in the data so significant that the models trained on this data could generalize to the real-world data with no additional training~\cite{tobin2017domain}. 
Thus, to address this \system includes a method for generating synthetic data streams by randomly varying over (1) tuple width (\# data items in a single tuple of a data stream), (2) its data types (the data type per data item), and (3) event rates (\# event tuples produced per time unit) crucial both for rigorously testing~\acrshort{sut}'s capabilities and collecting meaningful training data for ML models (cf. \Cref{tab:pspbench_evaluation_parameters} for data ranges).
While we generate data using the defined range, these values are highly configurable in the \system.  
To enable data streams from real-world applications, we use \texttt{kafka} as a data producer that is connected via \system to the \acrshort{sut}. 
We \textit{repeat} the data stream read from the source to mimic infinite data streams. 
\vspace{-3.0ex}
\begin{figure}
    \centering
        \includegraphics[width=0.7\linewidth]{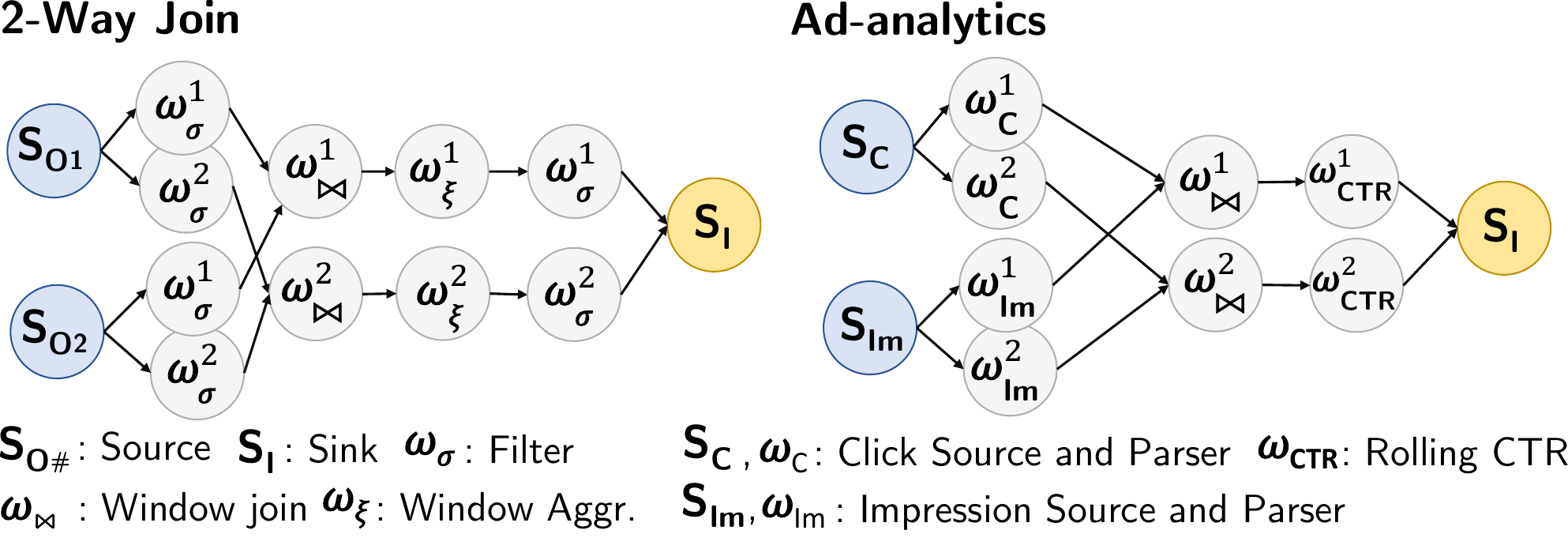}
        \label{fig:sub1}
        \vspace{-2ex}
    \caption{Example of \acrshort{pqp}: synthetic 2-way join and real-world ad analytics application.}
    \label{fig:pspbench_sp_ud_operators}
\end{figure}

\textit{\textbf{Query:}}  
%The strategic formulation and selection of \acrshort{pqp} is akin to designing a comprehensive benchmarking suite to assess various facets of~\acrshort{sps}’ performance, scalability, and resource utilization under varying workloads. 
For synthetic query plans, we offer an extensive range of~\acrshort{pqp} from an array of query structures, including 
simple linear queries with one filter to complex configurations involving multi-way joins and multiple chained filters. 
To give an example representation of such a 2-way join see \Cref{fig:pspbench_sp_ud_operators} (left). 
Moreover, we randomly enumerate over multiple parameters of these query structures such as filter function (e.g., $<$, $\leq$), its data type (the filter value's data type), window type (e.g., sliding, tumbling), window policy (e.g., time, count), etc., to generate queries that again can be used to evaluate a given \acrshort{sut} and is representative of workloads required to train a ML model. 
While we generate these query parameters randomly, an important question arises how we balance query properties, e.g., selectivity. 
For instance, random selection of filter literals may result that data never passes the generated filter. 
To avoid this, we use selectivity estimation methods~\cite{PrAg_ZeroTune_ICDE_2024} to estimate selectivity of given filter operator such that queries with only valid literals are generated where $sel_{\omega_\sigma} \text{is not } 0$.  
For real-world queries, we enable users to choose from the given range of applications available in our benchmark suite but also use them as a basis \acrshort{pqp} to generate more queries. 
For instance, consider the ad analytics application in \Cref{fig:pspbench_sp_ud_operators} (right) where the users can choose to generate more queries by adding more filter operators, choosing a different window count for the join, etc. 
This way, we allow users to execute given applications but also generate queries to evaluate \acrshort{sut} capabilities for dynamics and uncertainty inherent to real-world applications. 
Moreover, this flexibility allows to generation of representative \acrshort{pqp} aligned to the real world to train ML models. 
The full list of query applications is described in \Cref{tab:realWorldApplications} and explained in the next section. 
Other data ranges to configure query plans available in \system can be seen in \Cref{tab:pspbench_evaluation_parameters}.
%\todo{explain real world plans}
%\todo{we do not cover resource diversity}
\vspace{-0.5ex}

\textit{\textbf{Parallelism enumerator:}} 
While random enumeration is meaningful as it represents real-world data ranges for the parameters discussed so far, e.g., tuple widths for data stream and window length for operators, we note that random enumeration of parallelism degrees for operators is not the same. 
The random selection of parallelism degrees in \acrshort{pqp} will result in very noisy queries or even invalid queries, e.g., selecting higher parallelism degrees for downstream operators is less meaningful since there are anyways less tuples that have to be processed as tuples move down in the data flow graph (e.g., after filter operator). 
Moreover, random selection of parallelism degrees, e.g., $\omega_{\sigma} = 1$, $\omega_{\Join} = 10$ in the 2-way join example query in \Cref{fig:pspbench_sp_ud_operators} leads to a plan that is very bad in performance because it first limits processing capabilities by selecting only one instance of filter and hence there is limited use of $10$ instances of join operators and highly wasteful of resources. 
While such bad plans still might be interesting for benchmarking \acrshort{sut} to cover corner cases, learning ML models with such bad plans is not meaningful as they are not encountered in real-world. 
Thus, we employ different strategies for parallelism degree enumeration in \system that can be selected by the user depending on the needs. 
For instance, we provide \texttt{Rule-based} strategy that selects \textit{meaningful} parallelism degrees for \acrshort{pqp} derived based on literature~\cite{vasiliki2018Threesteps} but also random enumeration as defined below. 

\texttt{Random} selects a parallelism degree randomly within the given range, usually upto maximum number of cores available on physical resources, introducing variability for comprehensive performance assessment of \acrshort{sut}.
\texttt{Rule-based} goes beyond randomness and selects parallelism based on workload characteristics and physical resources. It considers factors such as event rates, operator selectivity, and the number of cores, enabling a more targeted enumeration of parallelism for upstream and downstream operators. This approach seeks to optimize performance by aligning parallelism with the specific demands and capacities of the system~\cite{vasiliki2018Threesteps}. 
\texttt{Exhaustive} aims to test every unique combination of parallelism degrees, ensuring that each combination is tested. 
%While comprehensive, its applicability is limited by the exponential increase in the number of combinations with the addition of more operators or parallelism degrees, leading to a potentially unmanageable number of queries. 
\texttt{MinAvgMax} cycles through generating queries with minimum, average, and maximum numbers of parallelism degrees, systematically exploring the effects of varying parallelism degrees on system performance, from least to most intensive use of resources. 
\texttt{Increasing} evaluates the impact of incremental change in parallelism, starting at the minimum degree and increasing stepwise to the maximum for each operator up the dataflow graph.
%Ideal for evaluating the impact of incremental changes in parallelism, this strategy starts with the minimum degree of parallelism and increases it incrementally for each operator until the maximum degree is reached. 
\texttt{Parameter-based} is designed for rapid testing, as it configures parallelism based on user input. % uniformly across operators, excluding source and sink, ensuring consistent and simplistic evaluations.
%Designed for scenarios needing quick testing or execution at a fixed parallelism degree, allows for parallelism configuration for all operators except the source and sink. The parallelism runtime parameter is used to set a uniform degree of parallelism, making it suitable for evaluations where consistency and simplicity are desired. 

\vspace{-3ex}
\subsection{Applications} \label{subsec:applications}
We include a selection of applications in the \system benchmarking suite by analyzing previous research works in databases~\cite{yahooChintapalli2016benchmarking,arasu2004linear} and stream processing~\cite{bordin2020dspbench,hesse2021espbench,Asterios2018benchmark,garcia2023spbench}. 
The applications are chosen based on a set of criterion that capture the diversity of streaming workloads, including data sources' tuple width, the data items'  type, as well as the different operators and their complexity, e.g., standard~\acrshort{sps} and user-defined operators in data flow graphs. 
To thoroughly assess~\acrshort{pdsp} in heterogeneous environments, we classified these applications into real-world and synthetic categories. 
This approach ensures a detailed evaluation of the~\acrshort{sut} capabilities, with a special emphasis on its performance (latency and throughput) but also readiness for future demands across various conditions from typical to peak usage scenarios for ML integration. 

The real-world applications reflect genuine data streams, such as \textit{social media feeds}, \textit{financial transactions} and \textit{IoT sensor data}, which are crucial for mimicking actual system loads and behaviors for the benchmarking process as presented in~\Cref{tab:realWorldApplications}. 
For instance, DEBS 2014 Smart Grid data~\cite{shukla2017riotbench} serves as a real-world benchmark, reflecting energy usage patterns from smart plugs. 
%These scenarios are critical as they offer authentic benchmarks that mimic actual system loads and user behaviours.
On the other hand, synthetic applications also represent real-world scenarios by including standard \acrshort{sps} operators like filters, aggregates and joins, but the data streams are generated artificially, allowing to stress test \acrshort{sut} under hypothetical future scenarios with high data volumes. 
This dual approach ensures a balanced assessment of~\acrshort{sut}'s performance, scalability, and adaptability, preparing it for current and future data processing challenges.
We enlist all the (both synthetic and real-world) applications included in \system in \Cref{tab:pspbench_evaluation_parameters}. 
While we provide the applications described above, \system can be easily extended by integrating new jobs from other benchmarks like YSB~\cite{yahooChintapalli2016benchmarking} and Nexmark~\cite{tucker2008nexmark}. 

%\vspace{-3ex}
\begin{table}[h]
\tiny
\centering
\begin{tabular}{lll}
\toprule
\textbf{Applications}                                                    & \textbf{Area}                                                   & \textbf{Description}                                                                                                                                                                                                                                                                                                                                  \\ \hline
Word Count (WC)~\cite{kulkarni2015Heron}                                                          & \begin{tabular}[c]{@{}l@{}}Text \\ Processing\end{tabular}      & \begin{tabular}[c]{@{}l@{}}Processes a text stream, tokenizes sentences into words, and counts the \\occurrences of each word in real-time using a key-based aggregation.\end{tabular}                                                                                                                                                                   \\ \hline
Machine Outlier (MO)~\cite{YxJiangMachineOutlier}                                                     & \begin{tabular}[c]{@{}l@{}}Network \\ Monitoring\end{tabular}   & \begin{tabular}[c]{@{}l@{}}Detects network anomalies in machine usage data streams using the\\ \textit{BFPRT algorithm}~\cite{yoon2007approach} to identify outliers based on statistical medians.
\end{tabular}                                                                                                                                     \\ \hline
Linear Road (LR)~\cite{arasu2004linear}                                                         & \begin{tabular}[c]{@{}l@{}}Traffic \\ Management\end{tabular}   & \begin{tabular}[c]{@{}l@{}}Processes vehicle-generated location data through four queries: \textit{toll} \\\textit{notification}, \textit{accident notification}, \textit{daily expenditure}, and \textit{total travel} \\ \textit{time}, to calculate charges or detect incidents.
\end{tabular}                                                                                                          \\\hline
\begin{tabular}[c]{@{}l@{}}Logs \\Processing (LP)~\cite{DomenicoSolazzoLogProcessing}\end{tabular}          & \begin{tabular}[c]{@{}l@{}}Web \\ Analytics\end{tabular}        & \begin{tabular}[c]{@{}l@{}}Processes HTTP Web Server log data to extract insights using two queries:\\ (i) counts visits within specified intervals, and (ii) tallies status codes. \end{tabular}                                                \\\hline
\begin{tabular}[c]{@{}l@{}}Google Cloud \\Monitoring (GCM)~\cite{karimov2018benchmarking}\end{tabular} & \begin{tabular}[c]{@{}l@{}}Cloud \\ Infrastructure\end{tabular} & \begin{tabular}[c]{@{}l@{}}Analyzes cloud computing data by calculating average CPU usage over\\ time, either grouped by job or category, with results processed through\\ sliding windows and specific grouping operators.\end{tabular}                                                   \\\hline
TPC-H (TPCH)~\cite{boncz2013tpc}                                                            & E-commerce                                                      & \begin{tabular}[c]{@{}l@{}}Processes a stream of order events to emit high-priority orders, utilizing\\ operators to structure, filter, and calculate the occurrence sums of order\\ priorities within specified time windows.\end{tabular}                                                                                                         \\\hline
\begin{tabular}[c]{@{}l@{}}Bargain \\Index (BI)~\cite{biem2010ibm}\end{tabular}            & Finance                                                         & \begin{tabular}[c]{@{}l@{}}Analyzes stock quotes streams to identify bargains by calculating the\\price-to-volume ratio against a threshold using \textit{VWAP} and \textit{Bargain} \\\textit{Index} Calculators, emitting qualifying quotes.\end{tabular} 
\\\hline
\begin{tabular}[c]{@{}l@{}}Sentiment \\ Analysis (SA)~\cite{SaurabhDubeySentimentAnalysis}\end{tabular}       & \begin{tabular}[c]{@{}l@{}}Social \\ Network\end{tabular}       & \begin{tabular}[c]{@{}l@{}}Determines the emotional tone of tweets by assessing sentiment using \\ TwitterAnalyzer and SentimentClassifier operators, which apply Basic \\or LingPipe classifiers to score and label the tweets.\end{tabular}                                                                                                  \\\hline
\begin{tabular}[c]{@{}l@{}}Smart \\ Grid (SG)~\cite{DEBS2014SmartGrid}\end{tabular}               & \begin{tabular}[c]{@{}l@{}}Sensor \\ Network\end{tabular}       & \begin{tabular}[c]{@{}l@{}}Analyzes smart home energy usage through two queries that calculate \\global and local average loads using sliding window.\end{tabular}                                                                                                                                                                                    \\\hline
\begin{tabular}[c]{@{}l@{}}Click \\ Analytics (CA)~\cite{lu2014streamBench}\end{tabular}          & \begin{tabular}[c]{@{}l@{}}Web \\ Analytics\end{tabular}        & \begin{tabular}[c]{@{}l@{}}Analyzes user interactions with online content through two queries:\\ grouping click events by Client ID for repeat and total visits per URL, \\and identifying geographical origins using a Geo-IP database.\end{tabular}                                                     \\ \hline
\begin{tabular}[c]{@{}l@{}}Spike \\ Detection (SD)~\cite{simmhan2011adaptive}\end{tabular}          & \begin{tabular}[c]{@{}l@{}}Sensor \\ Network\end{tabular}       & \begin{tabular}[c]{@{}l@{}}Processes sensor data streams from a production plant to detect sudden\\ temperature spikes by calculating average temperatures over sliding \\windows, and identify spikes exceeding 3\% of the average.\end{tabular}                                                                                               \\ \hline
\begin{tabular}[c]{@{}l@{}}Trending \\ Topics (TT)~\cite{mathioudakis2010twittermonitor}\end{tabular}          & \begin{tabular}[c]{@{}l@{}}Social \\ Network\end{tabular}       & \begin{tabular}[c]{@{}l@{}}Processes stream of tweets using the TwitterParser and TopicExtractor\\ operators to identify trending topics on Twitter based on aggregated \\popular topics based on predefined thresholds.\end{tabular}                                                                                   \\ \hline
\begin{tabular}[c]{@{}l@{}}Traffic \\ Monitoring (TM)~\cite{GeoTools}\end{tabular}       & \begin{tabular}[c]{@{}l@{}}Sensor \\ Network\end{tabular}       & \begin{tabular}[c]{@{}l@{}}Processes streaming vehicle data using TrafficEventParser and \\RoadMatcher operators to match vehicle locations to road segments then \\calculates average speed per segment using the AverageSpeedCalculator.\end{tabular}                                                               \\ \hline
Ad Analytics (AD)~\cite{neumeyer2010s4}                                                       & Advertising                                                     & \begin{tabular}[c]{@{}l@{}}Processes real-time data on user engagement with digital ads by parsing\\ clicks and impressions, calculating their counts within time windows, \\and computing the click-through rate (CTR) with a rolling CTR operator.\end{tabular}                                                             \\ \hline
Synthetic Queries                                                       & \begin{tabular}[c]{@{}l@{}}Standard DSP\\Queries\end{tabular}                                                     & \begin{tabular}[c]{@{}l@{}}Assess standard streaming workloads by randomly generating diverse\\data streams and query structures with increasing complexity. It supports \\various data types and standard operators like filter, window aggregate,\\window join, and groupby to evaluate streaming operators through \\synthetic query structures, from simple linear to complex multi-join queries.
\end{tabular}                                                             \\ \bottomrule
\end{tabular}
\caption{Benchmarked parallel query structures based on synthetic and real-world applications (based on~\cite{ihde2022survey,bordin2020dspbench,hesse2021espbench,garcia2023spbench,zeuch2019analyzing}.})
\label{tab:realWorldApplications}
\end{table}
%\vspace{-8ex}

%\todo{cite and refer open source}
%\todo{add a summary of all the applications and their description in a table?}
%\todo{add a section on ML models, explain the cost models and say that you later evaluate them.}
%enabling the generation of extreme test cases. 
%This aspect is vital for stress-testing the~\acrshort{sut}, evaluating its performance under peak loads, and scalability in response to high-volume data streams that mimic potential future real-world scenarios.

\vspace{-3ex}
\section{Evaluation}\label{sec:results}
In this section, we present an extensive evaluation using \system.  % evaluation 
Due to many workloads both synthetic and real-world and varying parameters in \system, we select interesting observations (\textbf{O\#}) from our evaluation for the following evaluation questions.

\textbf{Exp. 1: Impact of PQP complexity on performance.} How increasing parallel query complexity such as number and type of operators as well as parallelism degree can influence performance?

\textbf{Exp. 2: Impact of heterogeneous hardware on performance.} How heterogeneous hardware can impact the execution of \acrshort{pqp}?

\textbf{Exp. 3: Integration of ML models in \system.}(1) How different learned cost models perform on various \acrshort{pqp}? %Can learned cost models for streaming queries be evaluated using \system?}
(2) What is the influence of parallelism enumeration strategies on training efficiency?\\

\vspace{-2ex}
\textit{\textbf{Environment Setup and Implementation.}} We use the \textit{Cloudlab research testbed}~\cite{duplyakin2019design} to perform all our experiments as it provides the necessary distributed infrastructure %with both homogeneous and heterogeneous hardware
(cf.~\Cref{tab:pspbench_hardwares}) for configuring and deploying an \acrshort{sut} cluster, enabling us to benchmark using \system. 
For this initial evaluation, we select \textit{Apache Flink v1.16.1} as \acrshort{sut}, however this can be exchanged by any~\acrshort{sps}. 
In addition, \system uses an \textit{Apache Kafka} on a separate machine from \acrshort{sut} to produce data at different event rates for the various applications. %using \textit{input topic} followed by consuming the output provided \acrshort{sut} using \textit{output topic}. 
\acrshort{pqp} from different query structures ($\approx30k$~\acrshort{pqp}s) are executed three times for $3$ minutes each on clusters of $10$ nodes (cf.~\Cref{tab:pspbench_hardwares}) and corresponding performance metrics are collected and stored locally as well as in \textit{MongoDB} database. 
%We collected total $~30k$~\acrshort{pqp} for both synthetic ($\approx20k$) and real-world ($\approx10k$) applications using~\system.
Furthermore, the \textit{controller} is implemented in \textit{Django} and \textit{\acrshort{wui}} is developed with \textit{Vue.js} to take users' input, such as cluster setup, \acrshort{sut} deployment and~\acrshort{pqp} as described in~\Cref{sec:systemOverview}.

\textit{\textbf{Metrics.}} 
For experiments 1 and 2, we focus on \textit{end-to-end latency}, though \system can be used to measure other performance metrics depending upon \acrshort{sut} benchmarking requirements.  
\textit{End-to-end latency} represents the time interval starting from the production of the first data tuple from the data source until when the output of the query result is delivered to the data sink.  
It is the sum of the processing latency of each operator (including window time) in the processing pipeline, the network latency of data transmission from a data source within operators to the data sink, and the input and output latency of reading and writing data to and from external systems like IoT data sources. Given that operators might be distributed across different locations based on the CloudLab resources, network latency is a significant factor.~\cite{PrAg_ZeroTune_ICDE_2024,luthra_tcep_elsevier_journal_2021}    
We report the mean of three runs of measuring median latency (\textit{50th percentile}). % and the tail (\textit{95th percentile}). 

In experiments 3 and 4, we report Q-error q(c, c') which is a well-known metric to measure the accuracy of ML models~\cite{leisOptimizer2015}. 
In the context of learned cost models it gives relative deviation of the true cost c (latency) with its prediction c'. 
In addition, we compare the proposed enumeration strategies to generate parallelism degrees (cf. \Cref{sec:pspBench_PQP_IDG}) in terms of accuracy with Q-error and training time of the models using these strategies. 
\vspace{-3.0ex}
\begin{table}[]
\tiny
\centering
\begin{tabular}{ccc}
\toprule
\textbf{Diversity}      & \textbf{Parameters}                                                      & \textbf{Parameter Data Range}                                                                                                                                                                                                                                                                                                                     \\ \hline
\multirow{10}{*}{Query} & \begin{tabular}[c]{@{}c@{}}Real-world query structures\end{tabular}   &  %\begin{tabular}[c]{@{}c@{}}Word Count (WC), Machine Outlier (MO), Linear Road (LR), \\ Logs Processing (LP), Google Cloud Monitoring (GCM), TPCH, \\ Bargain Index (BI), Sentiment Analysis (SA), Smart Grid (SG), \\ Click Analytics (CA), Spike Detection (SD), Trending Topic (TT), \\ Traffic Monitoring (TM), Ad Analytics (AD)\end{tabular} 
                        Refer ~\Cref{tab:realWorldApplications} for full list\\ \cline{2-3} 
                        & \begin{tabular}[c]{@{}c@{}}Synthetic query structures\end{tabular}    & \begin{tabular}[c]{@{}c@{}}Linear, 2-chained filter, 3-chained filter, 4-chained filter, \\ 2-way join, 3-way join, 4-way join, 5-way join, 6-way join\end{tabular}                                                                                                                                                                               \\ \cline{2-3} 
                        & \begin{tabular}[c]{@{}c@{}}Parallelism \\ degree categories\end{tabular} & \begin{tabular}[c]{@{}c@{}}$1 \leq$ XS $< 8$, $8 \leq$ S $< 16$, $16 \leq$ M $< 32$, \\ $32 \leq$ L $< 64 $, $64 \leq$ XL $< 128$, $128 \leq$ XXL\end{tabular}                                                                                                                   \\ \cline{2-3} 
                        & Window duration (ms)                                                     & \begin{tabular}[c]{@{}c@{}}50, 100, 150, 200, 250, 325, 750, 1k, 1.5k, 2k, 2.5k, 3k, 4k, \\5k, 6k, 7k, 8k, 9k, 10k\end{tabular}                                                                                                                                                                                                                  \\ \cline{2-3} 
                        & Window length (tuples)                                                   & \begin{tabular}[c]{@{}c@{}}2, 3, 4, 5, 7, 10, 17, 25, 37, 50, 62, 75, 82, 100, 150, \\200, 250, 300, 350, 400\end{tabular}                                                                                                                                                                                                                       \\ \cline{2-3} 
                        & Sliding length (ratio)                                                   & {[}0.3, 0.4, 0.5, 0.6, 0.7{]} x Window length                                                                                                                                                                                                                                                                                                     \\ \cline{2-3} 
                        & Window types and policy                                                  & type: sliding and tumbling, policy: count and time-based                                                                                                                                                                                                                                                                                          \\ \cline{2-3} 
                        & Window aggr. functions                                                   & min, max, avg, mean, sum                                                                                                                                                                                                                                                                                                                          \\ \cline{2-3} 
                        & Join and filter data types                                               & string, integer, double                                                                                                                                                                                                                                                                                                                           \\ \cline{2-3} 
                        & Filter functions                                                         & \begin{tabular}[c]{@{}c@{}}$\leq, \geq, \neq, =, <, >$,\\ startsWith, endsWith, endsNotWith, startsNotWith\end{tabular}                                                                                                                                                                                    \\ \hline
\multirow{3}{*}{Data}   & \begin{tabular}[c]{@{}c@{}}Tuple width and data type\end{tabular}     & {[}1 - 15{]} x {[}str., doubles, int{]}                                                                                                                                                                                                                                                                                                           \\ \cline{2-3} 
                        & Event rate (events/sec)                                                  & 10, 100, 1k, 5k, 10k, 50k, 100k, 200k, 500k, 1mn, 2mn, 4mn                                                                                                                                                                                                                                                                                                  \\ \cline{2-3} 
                        & Partitioning strategy                                                 & Strategy for data distribution (forward, rebalance, hashing)                                                                                                                                                                                                                                                                                                                                                                                                                                                       \\ \hline
Resource                & Cluster type                                                             & Homogeneous: m510, Heterogeneous: c6320, c6525\_25g                                                                                                                                                                                                                                                                                                \\\hline 
ML models                & Learned cost models                                                             & \begin{tabular}[c]{@{}c@{}}Linear regression (LR)~\cite{ganapathiFlatVector2009}, Multi-layer perceptron (MLP)~\cite{hosseinzadeh2014multilayer}, \\Random forest (RF)~\cite{chen2016parallel}, Graph neural networks (GNN)~\cite{wu2024stage,PrAg_ZeroTune_ICDE_2024,heinrich2024costream}  \end{tabular}                                                                                                                                                                                                                                                                                          \\  
\bottomrule
\end{tabular}
\caption{\system benchmark parameters for \acrshort{sut} (order by complexity).}
\vspace{-8ex}
\label{tab:pspbench_evaluation_parameters}
\end{table}
\vspace{-7.0ex}
\begin{table}[]
\tiny
\centering
\begin{tabular}{ccccccccc}
\toprule
\textbf{\begin{tabular}[c]{@{}c@{}}Cluster Type\end{tabular}} & \textbf{Clusters} & \textbf{Node} & \textbf{CPU} & \textbf{\begin{tabular}[c]{@{}c@{}}RAM (GB)\end{tabular}} & \textbf{\begin{tabular}[c]{@{}c@{}}Disk (GB)\end{tabular}} & \textbf{Processor} & \textbf{\begin{tabular}[c]{@{}c@{}}Speed (Ghz)\end{tabular}} & \textbf{\begin{tabular}[c]{@{}c@{}}Network \\Link Speed\end{tabular}} \\ \hline
Ho                                                               & m510              & 10           & 8            & 64                                                           & 256                                                           & Xeon D             & 2                                                               & \multirow{3}{*}{\begin{tabular}[c]{@{}c@{}}10 \\ Gbps\end{tabular}}   \\
He                                                               & c6525\_25g        & 10           & 16           & 128                                                          & 480                                                           & AMD EPYC           & 2.2                                                             &                                                                       \\
He                                                               & c6320             & 10            & 28           & 256                                                          & 1024                                                          & Haswell            & 2.0                                                             &                                                                       \\
\bottomrule
\end{tabular}
\caption{Cluster of resources utilized on CloudLab tested to perform benchmarking to SUT. Here, “Ho” and “He” are homogeneous and
heterogeneous}
%\vspace{-2ex}
\label{tab:pspbench_hardwares}
\end{table}

\begin{figure*}[ht]
  \centering
  \begin{subfigure}[b]{0.8\textwidth}
    \includegraphics[width=\textwidth]{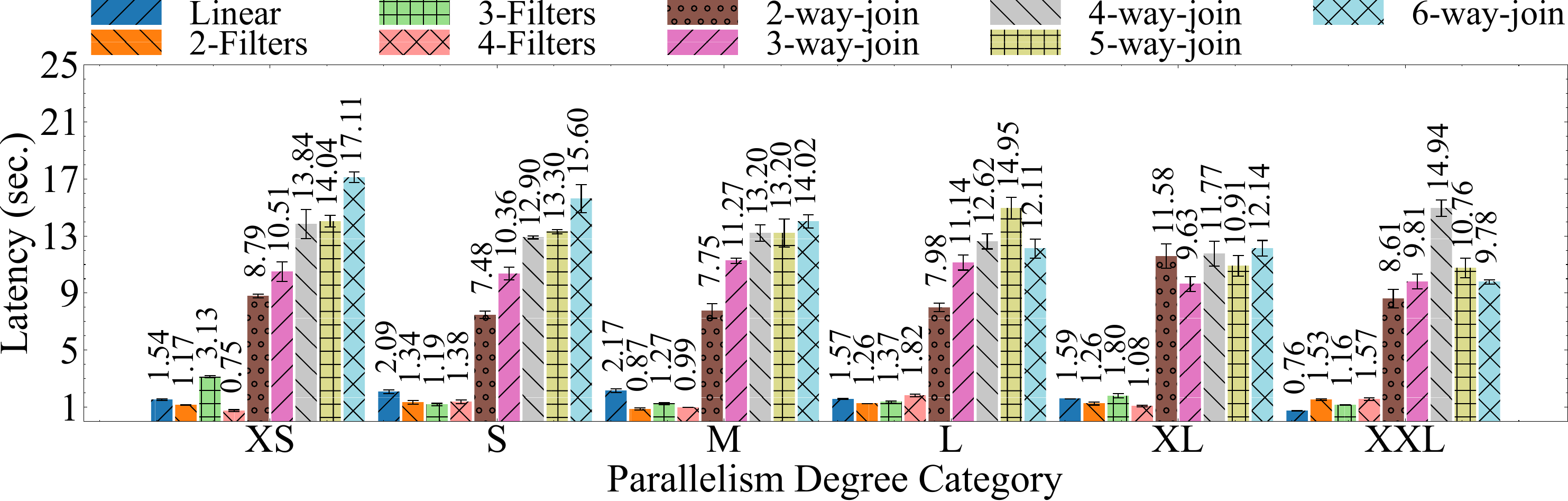}
    %\caption{Synthetic applications}
    %\label{fig:syntheticPQP_parallelismCategory}
  \end{subfigure}
  \hfill % optional; add some horizontal spacing
  \begin{subfigure}[b]{0.8\textwidth}
    \includegraphics[width=\textwidth]{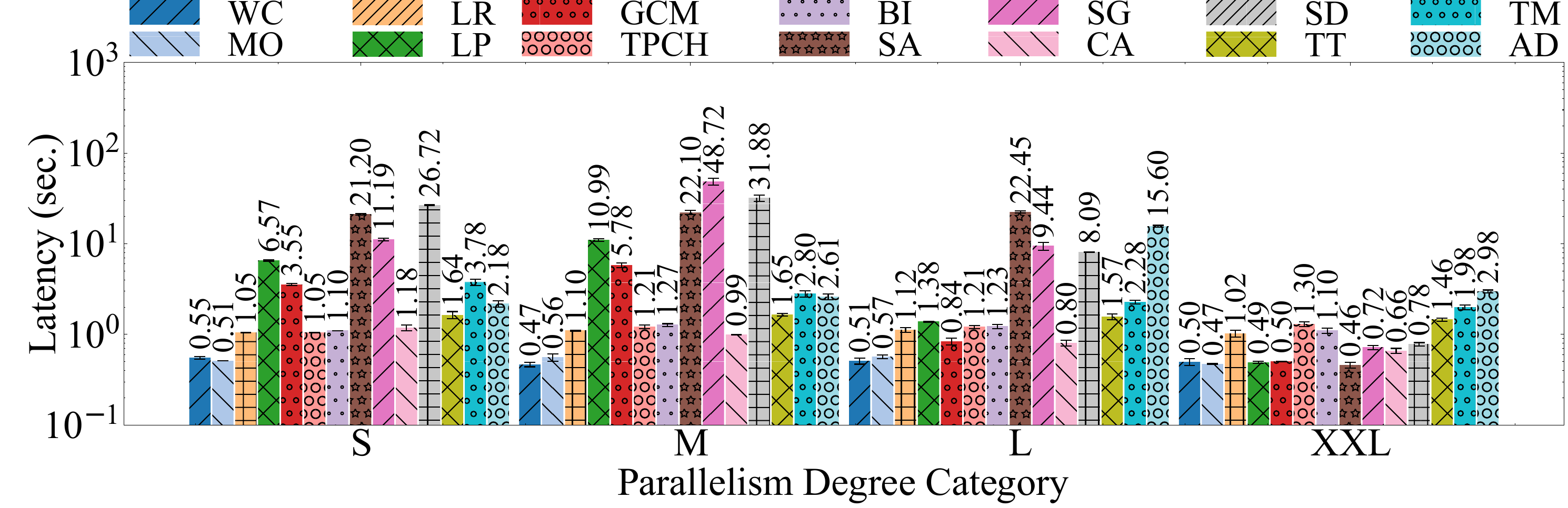}
    %\caption{Real-world applications}
    %\label{fig:realWorldPQP_parallelismCategory}
  \end{subfigure}
     \vspace{-2.0ex}
  \caption{Impact of parallelism degree on~\acrshort{pqp} performance from synthetic (top) and real-world applications (bottom). The analysis shows distinct performance behaviors, highlighting the benefits of parallelism in improving end-to-end latency and the need to consider query and operator characteristics in PDSP environments. Results indicate a complex interplay between standard operations and~\acrshort{udo}, with \textit{paradoxical} effects and \textit{non-linear} performance trends related to parallelism degree. (Note: Bottom figure omits XS and XL; their performance mirrors S and L)}%Note that we skip XS and XL as they have the same performance as S and L.}
  \label{fig:PQP_vs_parallelismCategory}
\end{figure*}
\vspace{-5.0ex}

\textit{\textbf{Evaluation Parameters.}}~\Cref{tab:pspbench_evaluation_parameters} outlines evaluation parameter range evaluated by \system for data stream, \acrshort{pqp} and resources (cf.~\Cref{subsec:workload_enum}). 
~\Cref{tab:pspbench_hardwares} presents the used hardware configuration from CloudLab testbed. 
Although we evaluate different event rates, we present results on the highest event rate of $4mn$ events/sec. unless otherwise specified as intuitively higher scale of events will benefit from parallelism. 
Thus, we consider the occurence of a number of events over a fixed interval of time, so data is modelled as poisson distributed since many real-world applications, e.g., network traffic, sensor networks, etc., are poisson distributed.
However, in \system we can also model other common data distributions such as zipf. 
%incorporating varying parallel query structures from both real-world and synthetic application,  to assess performance across different hardware configurations (see Table~\Cref{tab:pspbench_hardwares}) and with varying operator complexities. 
%Additionally, \system analyzes performance under diverse workloads, event rate and parallelism categories of query structures to explore a broad spectrum of~\acrshort{pdsp} scenarios. 
\vspace{-2.0ex}
\subsection{Exp. 1: Impact of PQP complexity}\label{subsec:unseenQueryPlans} 
We benchmark Flink using two~\acrshort{pqp} categories: a) synthetic, primarily with standard~\acrshort{sps} operators (cf.~\Cref{fig:PQP_vs_parallelismCategory} \textit{top}) and b) real-world, combining standard~\acrshort{sps} and~\acrfull{udo} (cf.~\Cref{fig:PQP_vs_parallelismCategory} \textit{bottom}). % as presented in the \Cref{tab:pspbench_evaluation_parameters}. 
%Although we evaluated different event rates, we present results on the highest event rate of $1mn$ events/sec. to analyze the effect of increasing parallelism degrees as intuitively higher scale of events will benefit from parallelism. 
We select homogeneous resources of \texttt{m510} cluster (with $10$ nodes) to analyze \textit{only} parallelism degree diversity. 
%To investigate the impact of increasing parallel query complexity on performance, especially on \textit{end-to-end latency}, we analyze both synthetic and real-world \acrshort{pqp} with different degrees of parallelism in \system. 
%\Cref{fig:PQP_vs_parallelismCategory} shows the performance of \acrshort{pqp}, each distinguished by its number and combination of standard~\acrshort{sps} operators and \acrfull{udo}, at an event rate of \textit{1M/second} on $m510$ clusters with $10$ nodes. 
Here, the complexity of a \acrshort{pqp} correlates both the composition of various operators and the parallelism degree applied to execute them. 
For instance, \textit{linear} parallel query structure, with a single source, multiple filters, and window aggregation without joins, has simpler data flow and lower computational demands, leading to reduced end-to-end latency.
%This simplicity typically leads to lower computational demands and, consequently, lower end-to-end latency. 
However, the presence of dual filters introduces computational requirements that can affect latency based on the data volume and filter complexity. 
%This simplicity posits lower computational overhead, suggesting an expectation of lower end-to-end latency. 
%However, even within this simplicity, the dual filters introduce a level of computational need that, depending on the data volume and filter complexity, can variably affect latency. 
%On the other hand, the complexity significantly increases with multi-way joins, starting from 2-way joins. 
%Each join operation, requiring the correlation of data from multiple sources, adds substantial computational and data management overhead, which requires parallel processing while managing the increased latency implications effectively. 
%We present interesting anecdotes of this analysis as follows. 
In contrast, multi-way joins significantly increase complexity and computational overhead, necessitating effective parallel processing to manage latency implications. 
We present key findings from this analysis.

\textit{\textbf{O1}- Increasing parallelism can speed-up multi-way join queries.} 
%In the context of~\system, the complexity cliff metaphorically refers to the point at which the complexity of managing~\acrshort{pqp} becomes significantly challenging. 
We observe an interesting trend in~\Cref{fig:PQP_vs_parallelismCategory} (top) when parallel query structures transitioned from linear query to more chained filters and joins. 
%Initially, when we add more filters, the latency remains almost consistent across parallelism categories \added{\textit{XS to XXL}, indicating minimal parallelism was sufficient.}  
%However, when we introduce join operators in~\acrshort{pqp}, it represents a tipping point where latency increases linearly, \added{reflecting the complexity of coordinating join operations across distributed datasets.}
Initially, adding filters keeps latency consistent across parallelism categories \textit{XS to XXL}. However, introducing join operators leads to a tipping point where latency increases linearly due to the complexity of coordinating joins across distributed datasets. 
At the same time, the parallel instances help in handling workload and reducing latency with increasing complexity. 
%A similar trend is also observed for \acrshort{pqp} from real-world applications in~\Cref{fig:PQP_vs_parallelismCategory} (bottom) where \acrshort{pqp} consists of combinations of varying numbers and types of standard~\acrshort{sps} operators and \acrshort{udo}. 
%For instance, applications with more standard~\acrshort{sp} operators such as \texttt{\textit{WC}}, \texttt{\textit{LR}} showed a consistent performance while \acrshort{pqp} with more computation intensive~\acrshort{udo} such as \textit{SA, SG, SD} has shown significant performance improvement with increasing parallelism degrees. 
%Thus, it highlights that parallelism can be beneficial for~\acrshort{pqp} with compute-intensive operators in significantly improving performance in comparison to~\acrshort{pqp} with complex structure but less computation-intensive operators. 
%\added{For instance,~\acrshort{pqp} with more standard~\acrshort{sps} operators such as \texttt{\textit{WC}}, \texttt{\textit{LR}} show consistent performance, while those with data-intensive~\acrshort{udo} such as \textit{SA, SG, SD} has shown significant performance improvement with increasing parallelism degrees.}
%\added{This highlights that parallelism benefits~\acrshort{pqp} with data-intensive operators more than those with less data-intensive operators.}
A similar trend is observed in real-world applications~\Cref{fig:PQP_vs_parallelismCategory} (bottom).~\acrshort{pqp} with standard~\acrshort{sp} operators such as \texttt{\textit{WC}}, \texttt{\textit{LR}} show consistent performance, while those with data-intensive \acrshort{udo} such as \texttt{\textit{SA, SG, SD}} show significant performance improvement with increasing parallelism.
This highlights that parallelism benefits~\acrshort{pqp} with data-intensive operators more than those with less data-intensive operators. 

\textit{\textbf{O2}- Increasing parallelism can speed-up queries, but not in all cases.} While increasing parallelism generally improves performance by distributing the workload, for some~\acrshort{pqp}, there is a paradoxical effect on end-to-end latency as~\acrshort{pqp} complexity increases. 
Beyond a certain threshold of parallelism (\textit{$32 \leq L <64$}), particularly with multiple joins, the overhead of managing parallel operations, such as data shuffling and synchronization, outweighs the benefits, leading to increased latency. 
%This scenario is comparable to traffic flow in a city where, beyond a certain volume, adding more lanes (parallel paths) can actually lead to more congestion due to bottlenecks at merge points and intersections. 
This is similar to adding more lanes in city traffic, which can cause congestion at merge points. 
Thus, performance improvements in multi-way joins are small or negligible as parallelism increases from $L$ to $XL$. 
%In contrast, the \acrshort{pqp} from real-world applications starts showing the benefit of parallelism on performance improvement in latency when the parallelism degree becomes extremely high.  
%For instance, with parallelism category \textit{$32 \leq L <64$}, \acrshort{pqp} from \textit{\texttt{SG}}, \texttt{\textit{SD}}, has shown significant improvement while \texttt{\textit{AD}} has shown negligible effect on performance even when the parallelism is more than $128$, showing the complexity of operators in the query. 
In contrast, \acrshort{pqp} from real-world applications show performance benefits at extremely high parallelism. 
For instance, \textit{$32 \leq L < 64$} parallelism significantly improves latency in \texttt{SG} and \texttt{SD}, while \texttt{AD} shows negligible performance gains even beyond \textit{128}, reflecting the complexity of operators in the query.  

\textit{\textbf{O3}- Queries with \acrshort{udo} shows unpredictable performance.} 
We also investigate the performance intricacies of standard~\acrshort{sp} operators compared to~\acrshort{udo}.%within~\acrshort{pdsp} environment. 
Our findings reveal distinct scalability and computational overhead that underline the relationship between operator types and parallel processing efficiency. 
Standard~\acrshort{sp} operators exhibit predictable scalability due to their well-defined semantics. For instance, a \textit{flatMap} operation in a \texttt{\textit{WC}} application scales almost linearly with increased parallelism, %as it can be independently executed across different partitions of the data without the need for complex state management or coordination. 
requiring no complex state management. 
%On the other hand, \acrshort{udo}, which allows for embedding of custom logic into the data processing pipeline, introduces variability in performance scalability, primarily due to the potential for complex state handling and the need for coordination across different execution units. 
%For instance, in the \texttt{\textit{AD}} application, the presence of multiple \acrshort{udo} performing custom aggregation and joining logic on a sliding window introduces a non-linear scaling behavior. 
%As parallelism increases, the overhead of managing state consistency and communication between parallel instances leads to a sublinear performance increase, and in some cases, performance may even degrade due to the overhead exceeding the computational benefits of parallelism. 
Conversely, \acrshort{udo}, which embed custom logic, show variable scalability due to state handling and coordination needs. In the \texttt{AD} application, custom aggregation and joining logic on a sliding window result in non-linear scaling, where increased parallelism leads to higher overhead, sometimes degrading performance.

\begin{figure*}[ht]
  \centering
  \begin{subfigure}[b]{0.8\textwidth}
    \includegraphics[width=\textwidth]{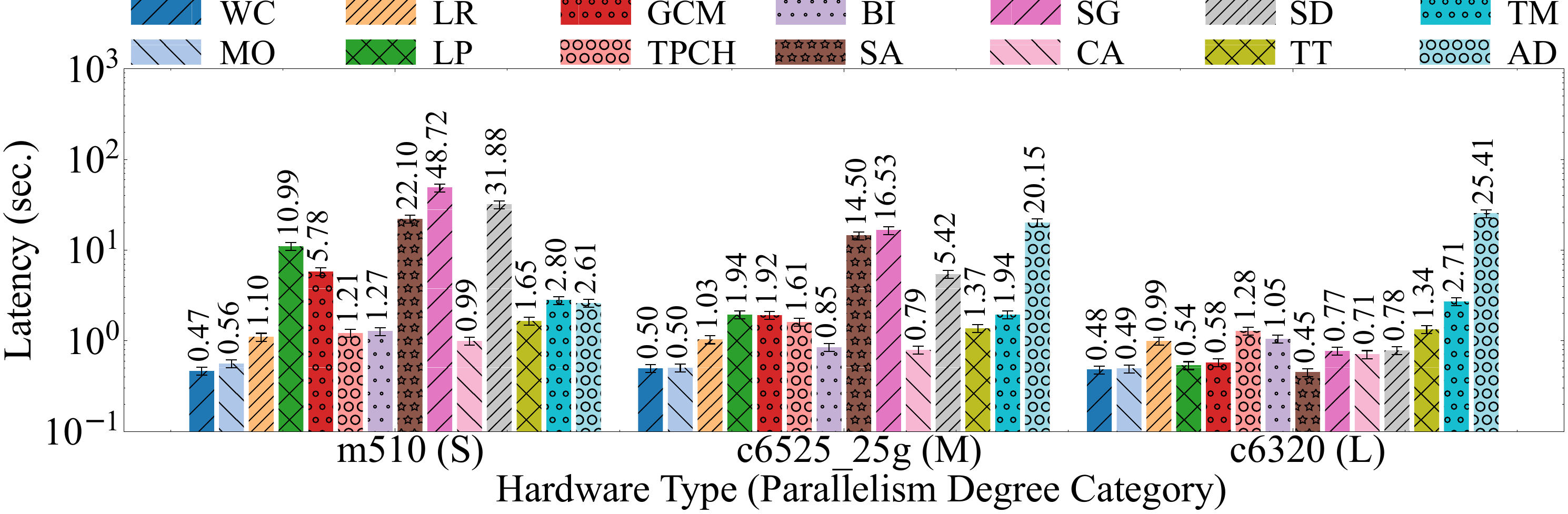}
    %\caption{Realworld application: impact on latency heterogeneous hardware}
    \label{fig:syntheticPQP_heterogeneity}
    \vspace{-2.0ex}
  \end{subfigure}
  \begin{subfigure}[b]{0.8\textwidth}
    \includegraphics[width=\textwidth]{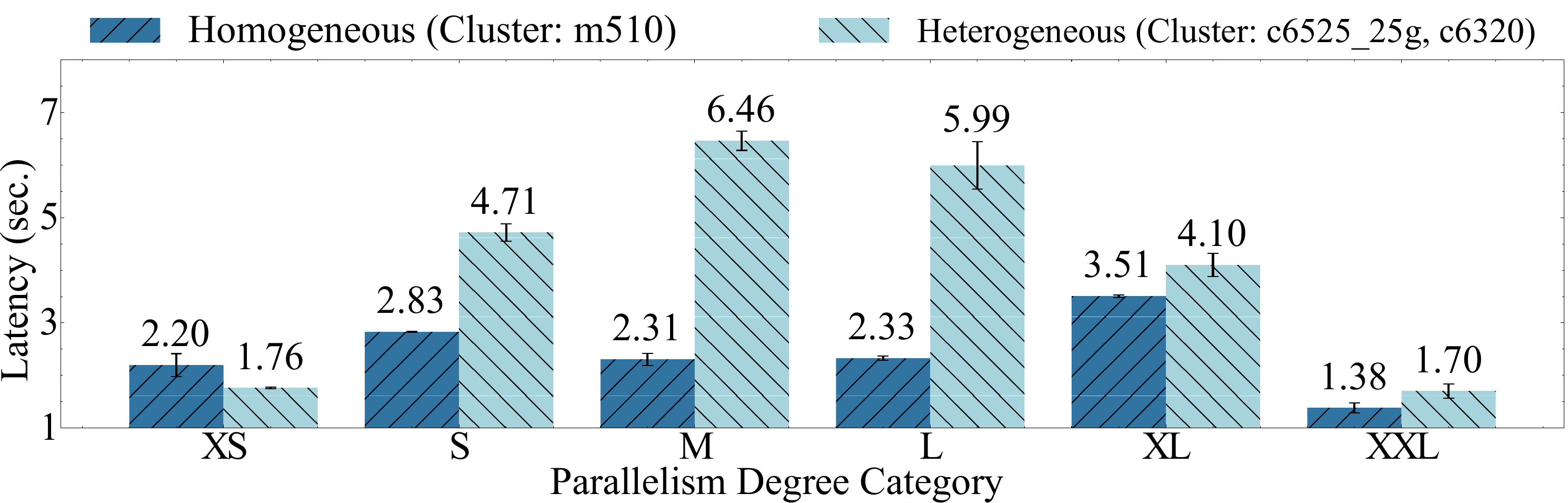}
    %\caption{Synthetic application: impact on latency heterogeneous hardware}
    \label{fig:realWorldPQP_heterogeneity}
  \end{subfigure}
    \vspace{-4.0ex}
  \caption{Impact of heterogeneous hardware on performance for~\acrshort{pdsp}, with varying parallelism degree and resource processing capabilities on real-world (top) and synthetic (bottom) applications. Evaluation shows that parallel processing benefits from hardware diversity but requires understanding hardware characteristics and workload distribution to optimize performance and avoid of pitfalls such as diversity dilemma.}
  \label{fig:PQP_vs_heterogeneity}
\end{figure*}
%\textit{\textbf{A4}- Balancing act:} These observations underscore the importance of understanding the trade-offs between the expressiveness of \acrshort{udo} and the predictable scalability of standard~\acrshort{sps} operators. While \acrshort{udo} provide the flexibility to implement complex processing logic, they also introduce challenges in achieving optimal performance in highly parallelized environments. 
%Consequently, when designing~\acrshort{pqp} for stream processing applications, it is crucial to balance the use of \acrshort{udo} with the inherent scalability characteristics of standard~\acrshort{sps} operators for efficient and predictable execution. 

\textit{\textbf{O4}- The non-linear effect of parallelism on latency:} \system provides a critical insight into the relationship between parallelism degree of \acrshort{pqp} and performance is non-linear. 
As the parallelism of standard~\acrshort{sp} operators in a query increases, the performance does not linearly increase as not necessarily each \acrshort{pqp} advantages from parallelism. 
This results from the previous observations O1 - O3 because of different factors like complexity of operators and paradox of parallelism. 
For instance, \textit{SA, SG, SD} has high latency for parallelism categories $S$ and $M$ which starts improving at level $L$, with significant improvements when parallelism exceeds $128$. 
% so does the challenge of optimizing end-to-end latency. 
%This analysis not only highlights the inherent trade-offs in~\acrshort{pdsp} systems but also underlines the importance of strategic query planning and execution optimization to mitigate the latency impacts of increased query complexity.
%In addition, \system provides empirical evidence on the impact of operator type on the scalability of parallel stream processing applications. 
%It highlights the need for careful consideration of operator characteristics in the design and optimization of \acrshort{sps}, particularly in the context of varying parallelism degrees.
\vspace{-2.0ex}
\subsection{Exp. 2: Impact of heterogeneous hardware on performance} 
%\system underscores a critical insight the relationship between query complexity and performance is non-linear and nuanced. As the number and type of relational algebra operators in a query increase, so does the challenge of optimizing end-to-end latency. This analysis not only highlights the inherent trade-offs in parallel and distributed stream processing systems but also underscores the importance of strategic query planning and execution optimization to mitigate the latency impacts of increased query complexity.              
Next, we evaluate the impact of homogeneous (\texttt{m510}) and heterogeneous hardware (\texttt{c6525\_25g,c6320}) clusters on performance for~\acrshort{pdsp} and executed~\acrshort{pqp} across clusters of $10$ nodes each. 
~\Cref{fig:PQP_vs_heterogeneity} (top) represents the mean \textit{end-to-end latency} of~\acrshort{pqp} with parallelism degree category as per \# cores on hardware of each cluster to analyze the performance of real-world applications. For instance, \texttt{m510} cluster has hardware with $8$ cores, so selected \acrshort{pqp} with $S$ parallelism degree category. Similarly, \acrshort{pqp} with parallelism degree categories $M$ and $L$ as clusters \texttt{c6525\_25g} and \texttt{c6320} have hardwares with $16$ and $28$ cores, respectively. 
~\Cref{fig:PQP_vs_heterogeneity} (bottom) shows the mean \textit{end-to-end latency} across different parallelism categories of~\acrshort{pqp} for three types of clusters for synthetic applications.  
%In~\Cref{fig:realWorldPQP_heterogeneity}, represents the maximum parallelism on each type of hardware based on the number of cores on the machines. 

\textit{\textbf{O5}- Powerful heterogeneous environment does not necessarily accelerate queries.} By evaluating~\acrshort{pdsp} across diverse hardware configurations, we encounter the \textit{diversity dilemma} where the theoretical advantages of heterogeneous environments sometimes clash with practical performance outcomes in~\Cref{fig:PQP_vs_heterogeneity} (top).
We anticipate that the diversity in computational processing capabilities would universally accelerate parallel processing and enhance performance. 
However, we notice that while applications \texttt{\textit{SA}}, \texttt{\textit{CA}}, \texttt{\textit{SD}} significantly benefited, % from the increasing processing power of underlying hardware and can achieve an 
showing exponential decrease in latency. 
On the other hand, \texttt{\textit{AD}} struggles to improve the performance in heterogeneous configuration due to the complexity of \acrshort{udo} coupled with the communication overhead across different instances. 
This finding suggests that the theoretical benefits of hardware diversity require careful orchestration for workload distribution and resource management strategies to leverage heterogeneous environments effectively. 

\textit{\textbf{O6}- Finding optimal parallelism for queries is non-trivial.} We also evaluate parallel processing capabilities of synthetic~\acrshort{pqp} on various hardware clusters as shown in~\Cref{fig:PQP_vs_heterogeneity} (bottom). 
We notice no consistent balancing point of parallelism exists for all workloads. Until this point, increased parallelism might not yield further benefits and could even hinder performance due to higher communication and synchronization overhead. Finding this point in a heterogeneous environment might be even more challenging. 
%For instance, as parallelism increases, we observe varying performance across both homogeneous and heterogeneous clusters. 
%\added{In the homogeneous clusters, we observe a trend where latency generally increases with parallelism from $XS$ to $XL$, then decreases at $XXL$.}  
%\added{A similar trend is seen in heterogeneous clusters, where latency is highest until the $M$ parallelism category and decreases for $L$, $XL$, and $XXL$.}
%\added{This suggests that the heterogeneous clusters benefit more from increased parallelism due to diverse computational capabilities.}
%The initial higher latency at $M$ parallelism
%\added{may result from workload distribution imbalance, which improves as parallelism grows, better utilizing diverse resources and redu cing overall latency.}
In homogeneous clusters, latency generally increases with parallelism from $XS$ to $XL$, then decreases at $XXL$. 
A similar trend is observed in heterogeneous clusters, with latency highest at $M$ and decreasing for $L$, $XL$, and $XXL$. 
This suggests heterogeneous clusters benefit more from increased parallelism due to diverse computational capabilities. Initial higher latency at $M$ may result from workload distribution imbalance, improving as parallelism grows and better utilizing diverse resources.

\textit{\textbf{O7}- Homogeneous or heterogeneous clusters: There is no clear choice for all cases.} \system provides insight about notable performance enhancement with increasing parallelism and diverse hardware configurations as presented in ~\Cref{fig:PQP_vs_heterogeneity} (top). 
Specifically,~\acrshort{pqp} from real-world applications benefit significantly from increasing parallelism and hardware processing capability, leading to improved performance. 
Conversely,~\acrshort{pqp} from synthetic applications demonstrate better performance on homogeneous clusters than heterogeneous ones, due to the efficient handling of standard~\acrshort{sp} operators.
%This pattern can be attributed to the nature of synthetic~\acrshort{pqp} encompass standard~\acrshort{sp} operators, which are more efficiently handled in homogeneous clusters. 
%Conversely, despite their potential for high computational power, heterogeneous clusters present challenges in achieving even workload distribution and increased communication and synchronization overheads due to varying communication speeds across different hardware units. 
Conversely, in heterogeneous clusters, despite high computational potential, challenges arise from uneven workload distribution and increased communication overheads due to varying speeds across different hardware units.

\vspace{-2.0ex}
\subsection{Exp. 3: Integration of ML models in \system} 

\textbf{(1) Performance of learned cost models.}
The \texttt{ML Manager} (cf.~\Cref{sec:systemOverview}) of \system offers benchmarking of performance of various ML models tailored for parallel and distributed stream processing environments. 
It uses data collected during benchmarking as labeled datasets for training and inference. % purposes which are collected during benchmarking performance. 
In this evaluation, we focus on assessing the effectiveness of different learned cost models in predicting performance of streaming queries
such as end to end latency.  
\vspace{-2.0ex}
\begin{figure*}[ht]
  \centering
    \includegraphics[width=0.75\textwidth]{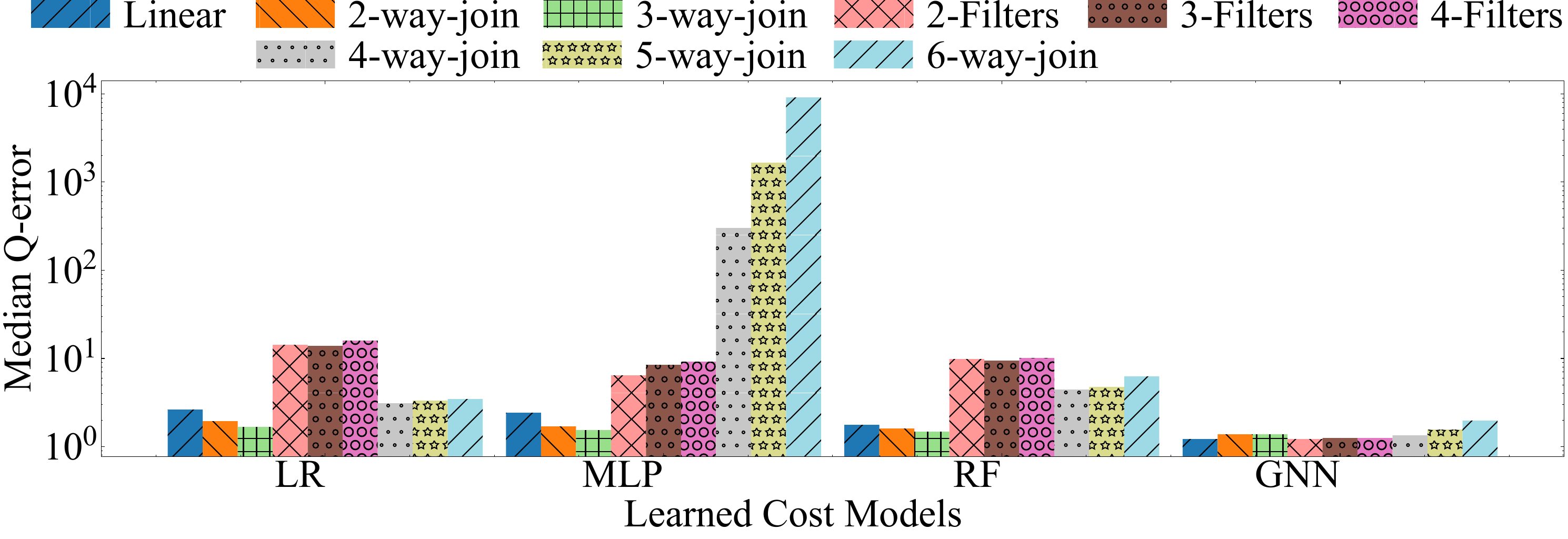}
    \vspace{-2.0ex}
  \caption{Comparison of accuracy of various ML models: Linear regression (LR)~\cite{ganapathiFlatVector2009}, Multi-layer perceptron (MLP)~\cite{hosseinzadeh2014multilayer}, Random forest (RF)~\cite{chen2016parallel}, Graph neural networks (GNN)~\cite{wu2024stage,PrAg_ZeroTune_ICDE_2024,heinrich2024costream} for various parallel query structures.}
  \label{fig:ml_models_integration}
  %\vspace{-2.0ex}
\end{figure*}

We integrate four distinct ML models architectures into the \texttt{ML Manager}: %Each model has been chosen for their potential in handling different aspects of performance prediction.
%These models include 
(1) Linear regression (LR)~\cite{ganapathiFlatVector2009}: traditionally used for its simplicity and effectiveness in prediction tasks, (2) Multi-layer Perceptrons (MLP)~\cite{hosseinzadeh2014multilayer}: known for capturing nonlinear relationships in data, (3) Random forest (RF)~\cite{chen2016parallel}: utilizes decision trees to improve prediction accuracy, and (4) Graph neural networks (GNN)~\cite{wu2024stage,PrAg_ZeroTune_ICDE_2024,heinrich2024costream}: applies graph structures to model complex relationships in data. It encodes~\acrshort{pqp} as a~\acrshort{dag}~\cite{PrAg_ZeroTune_ICDE_2024} within GNN, allowing the model to treat different operators within~\acrshort{pqp} as nodes, and the relationships between them as edges. 
These models are selected based on their diverse approaches to model and handle the complexities of stream processing queries. 
%Particularly noteworthy is GNN model, which encodes~\acrshort{pqp} as a~\acrshort{dag}~\cite{PrAg_ZeroTune_ICDE_2024}, 
%This representation is particularly suited to GNNs as it allowing the model to treat different operators within the PQP as nodes, and the relationships between them as edges.
The accuracy of these models is measured using Q-error $q(c, c')$ \textit{where $q \geq 1$} metric as mentioned before. 
Here, q-error being close to 1 represents better prediction accuracy.  
We also implement early stopping for each ML model to prevent overfitting and ensure efficient training times. 
Early stopping is based on monitoring the validation loss, halting training if it did not improve for $100$ consecutive epochs. 
This method was uniformly applied across all models to maintain consistency. 
The training of these model performed on $m510$ clusters. 
%Q-error is particularly well-suited for regression tasks where understanding the variability between predicted and actual costs is crucial. 

\textbf{O8-} \textit{Graph representation assists in learning dynamic behavior of SPS.} The primary aim of benchmarking ML models is to evaluate how these diverse cost models perform across various streaming queries. 
In~\Cref{fig:ml_models_integration} indicates that the GNN model consistently surpasses other models in predicting cost \ie latency, even as query complexity increases. The high accuracy of GNN can be attributed to the fact that the graph-based representation enables it to capture and utilize the intricate dependencies within the query structures effectively and dynamic behaviors of stream processing systems, leading to more accurate and reliable performance predictions compared to other models.
Similar observation was made for PQP related to real-world applications and thus we only present results on synthetic PQPs. 

\textbf{(2) Influence of ML training strategies.} 
%What is the influence of parallelism enumeration strategies on training efficiency?
\begin{figure}[t]
  \centering
  \begin{subfigure}[b]{0.55\textwidth}
    \centering
    \includegraphics[width=\linewidth]{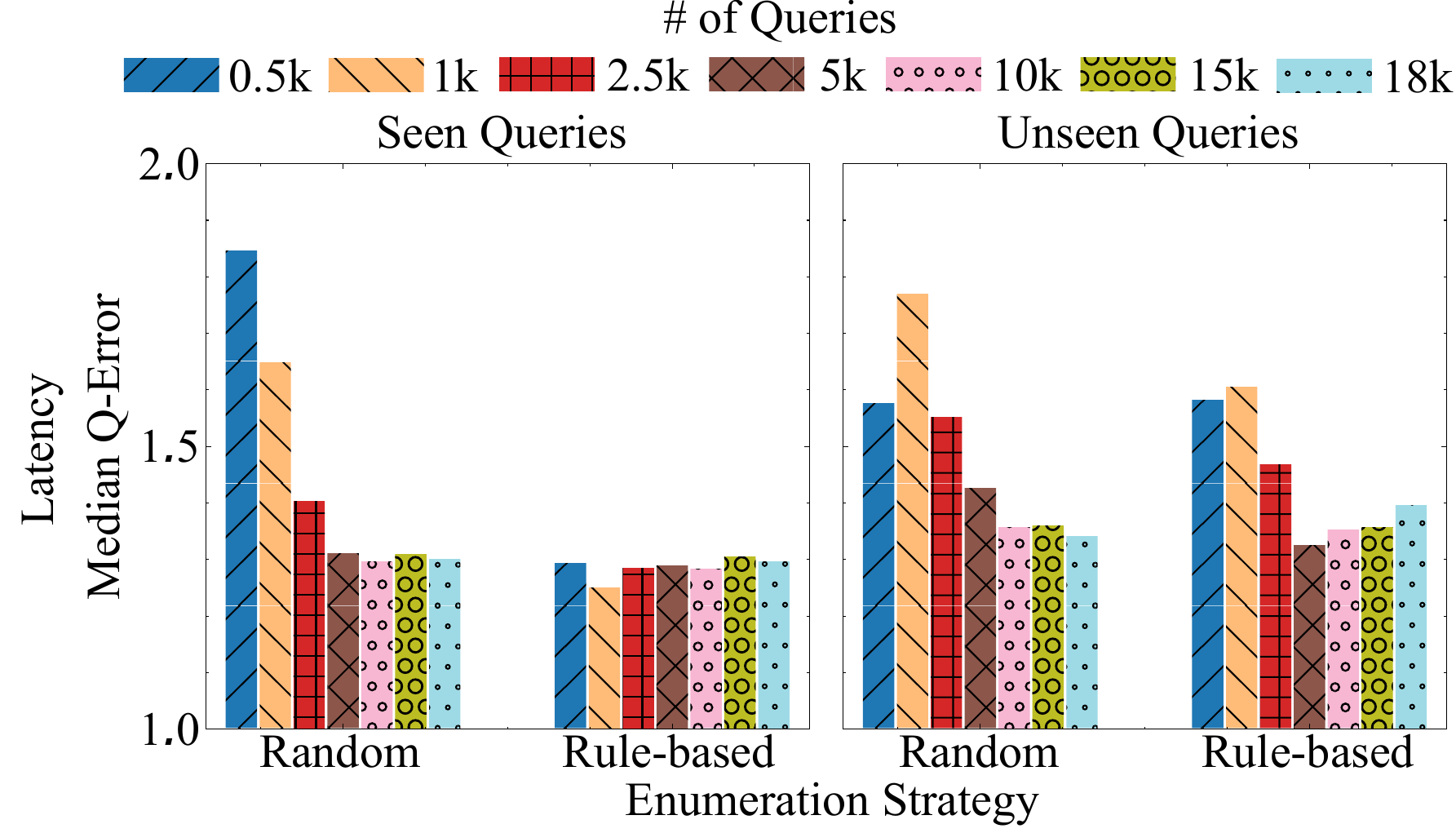}
    \vspace{-4.0ex}
    \caption{Amount of queries}
    \label{subfig:seen_enumeration_amount_of_queries}
  \end{subfigure}
 % \hspace{-1ex}
  \begin{subfigure}[b]{0.29\textwidth}
    \centering
    \includegraphics[width=\linewidth]{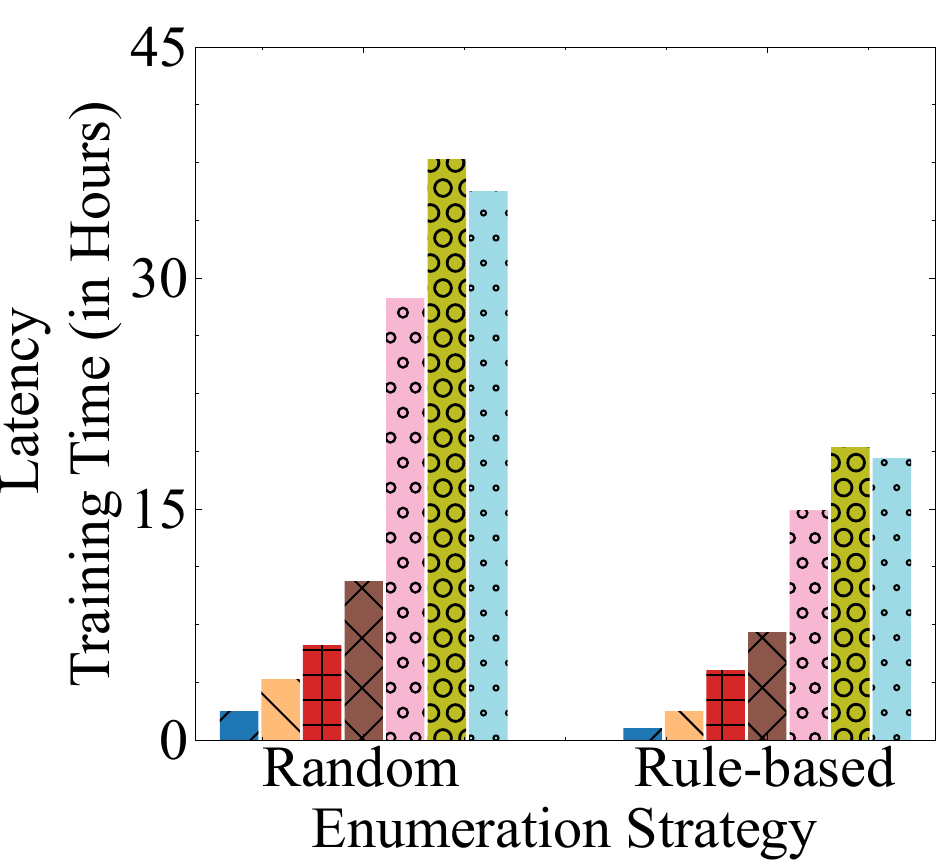}
    \vspace{-4.0ex}
    \caption{Training time}
    \label{subfig:training_time_vs_enumeration_strategy}
  \end{subfigure}
  \vspace{-2.0ex}
  \caption{Comparison of accuracy enumeration strategies. The rule-based strategy achieves better accuracy with (a) $2.5$k queries compared to the random strategy. It converges in (b) $4.5$ hours, three times faster than the random approach.}
  \vspace{-1ex}
  \label{fig:amount_of_queries_vs_enumeration_strategy}
\end{figure}
%In the context of~\acrshort{pdsp}, the dynamic nature of the environment poses significant challenges for data collection. 
%This complexity arises due to the variability in workload, resource configurations, and degrees of parallelism, which are critical factors that influence the training dataset's comprehensiveness. 
%Addressing these challenges requires substantial efforts in data collection and training, which are both time-intensive and resource-heavy.
%In \system, we explore various enumeration strategies (cf.~\Cref{subsec:workload_enum}) to optimize data collection efforts and reduce training time. 
We investigate the influence of various enumeration strategies (cf.~\Cref{subsec:workload_enum}), designed to optimize data collection efforts and reduce training time in \system. 
For instance, ~\Cref{fig:amount_of_queries_vs_enumeration_strategy} represent our findings by comparing the efficacy of random and rule-based parallelism enumeration strategies for training GNN models for latency prediction.
We stick to GNN models for this evaluation because of the observations made in the previous evaluation (O8).

\textbf{O9-} \textit{Data-efficient training for high accuracy with reduced training time.}~\Cref{subfig:seen_enumeration_amount_of_queries} illustrates that the GNN model trained using rule-based enumeration strategy requires only $2.5k$ queries for accurate latency predictions for both seen (linear, 2-way and 3-way join) and unseen (other remaining synthetic) queries. 
This efficiency is attributed to the strategy's systematic collection of data, which focuses on determining and exploring around selected parallelism degrees and generating more meaningful graph representations for GNN than the random strategy. 
%Such targeted data collection enables the GNN model to train more effectively in data-efficient manner, capturing intricate dependencies within the data-flow graphs more rapidly than the random strategy.
Further analysis in~\Cref{subfig:training_time_vs_enumeration_strategy} reveals that the rule-based strategy requires significantly less training time—approximately three time less than the time of the random strategy, \ie $4.6$ hours. 
%This stark contrast highlights the efficiency of rule-based enumeration in streamlining the training process.} 
%Our comprehensive benchmarking of enumeration strategies for data collection and model training underscores a crucial point: moving towards a data-efficient methodology not only conserves resources but also enhances the accuracy and speed of performance prediction models. Such strategies are vital for developing robust \acrshort{pdsp} systems capable of adapting to dynamic operational conditions with minimal overhead.
This benchmarking of enumeration strategies underscores a crucial insight: adopting data-efficient training methods using \system can significantly enhance model training efficacy. 
\vspace{-3ex}
\section{Related Work}
\label{sec:rwork}
\noindent
%   \todo{add hardware conscious benchmarks https://dl.acm.org/doi/10.14778/3303753.3303758}
We divide existing benchmarking systems for~\acrshort{dsp} systems into - \first DSPS- \ii TPC and \iii machine learning (ML) in benchmarking systems.

\textbf{Benchmarking Systems for DSPS.} Despite numerous benchmarks for database management systems, standardised and systematic benchmarks specifically designed for stream processing architectures are scarce. Our review of existing systems~\cite{ihde2022survey,ruben2019,liu2020resource,isah2019survey,hesseSurveyDSP2015,karimov2018benchmarking,boden2019benchmarking,boden2017benchmarking} reveals significant gaps, particularly in addressing the nuances of~\acrshort{dsp} (cf.~\Cref{tab:relatedWork}). 
The LRB~\cite{arasu2004linear} and similar efforts like the YSB~\cite{yahooChintapalli2016benchmarking,chintapalli2015benchmarking,grier2016extending} and BigDataBench~\cite{wang2014bigdatabench} %have expanded the scope of benchmarks but
often remain focused on batch processing rather than real-time streaming. Emerging micro-benchmarks such as HiBench~\cite{huang2010hibench}, StreamBench~\cite{wang2016stream}, RIoTBench~\cite{shukla2017riotbench} and OSPBench~\cite{van2020evaluationOSPBench} bring advancements in streaming benchmarks but still fall short in adequately testing scalability and handling real-time streaming requirements. 
These benchmarks typically do not address essential aspects of~\acrshort{dsp} systems like parallelism, hardware diversity, and variable workloads comprehensively. 
%Recent developments in benchmarking systems such as DSPBench~\cite{bordin2020dspbench} and SPBench~\cite{garcia2023spbench} have begun to address aspects of parallelism or concert processing. 
%Despite this, these benchmarks fail to present a holistic view of parallel stream processing. 
%These benchmarks often neglect key aspects like the degree of parallelism and data partitioning strategies, which are essential for benchmark \acrshort{dsp} system for optimal system performance and resource efficiency. Moreover, they also often rely on homogeneous hardware setups that do not reflect real-world~\acrshort{dsp} application environments.
%Additionally, there is a notable deficiency in automation support across benchmarks, which complicates benchmark deployment, operation, and result analysis, further highlighting the need for more sophisticated and automated benchmarking frameworks in the stream processing domain.} 
Recent benchmarking systems such as DSPBench~\cite{bordin2020dspbench} and SPBench~\cite{garcia2023spbench} address some aspects of parallelism but do not provide a holistic view of parallel stream processing.
They often neglect critical elements such as the degree of parallelism, data partitioning strategies, and typically rely on homogeneous hardware setups, which do not reflect real-world \acrshort{dsp} environments~\cite{zeuch2019analyzing}.
%Additionally, the lack of automation support complicates benchmark deployment, operation, and result analysis, highlighting the need for more sophisticated and automated benchmarking frameworks in the stream processing domain.

\textbf{TPC Benchmarks.} TPC~\cite{poess2000TPC} has developed several benchmarks to evaluate various aspects of computing systems~\cite{chen2011TPCDE,boncz2013TPCH,nambiar2006makingTPCDS,bond2016profilingTPCxV,taheri2019characterizingTPCxHci,cao2017bigbenchTPCxBB,poess2018analysisTPCxIoT}. 
Among these TPC benchmarks, TPCx-IoT~\cite{poess2018analysisTPCxIoT} is the most relevant to~\acrshort{dsp} as it addresses scenarios involving continuous data ingestion and real-time analytics.% which are core to stream processing.  
%This benchmark considers the performance of processing systems under the load of IoT-generated data streams, assessing aspects like throughput and latency, which are critical in stream environments. 
%While they offer valuable insights into system performance under a range of conditions, their applicability to the domain of parallel and distributed stream processing varies due to fundamental differences in system architecture and operational goals. 
%However, TPCx-IoT and other TPC benchmarks do not fully capture the specific challenges associated with parallel and distributed stream processing systems.
While the TPC benchmarks have contributed to standardizing the evaluation of transactional and analytical processing systems, there remains a distinct gap in benchmarking system designed specifically for parallel and distributed stream processing due to fundamental differences in system architecture and operational goals. 

TPC benchmarks provide extensive coverage of database and batch processing scenarios, which are more oriented towards singular query execution on static or slowly evolving data sets.
These benchmarks fall short in addressing the unique requirements of stream processing systems such real-time data processing, continuous ingestion, and immediate response to dynamic changes in data streams. 
%For instance, stream processing applications where multiple streams can be queried and correlated in real-time. 
In addition, none of the existing TPC benchmarks are designed to assess performance under these conditions, focusing instead on batch or transactional processing where data latency and continuous data flow are less critical. 
They do not adequately test data streaming partitioning and required parallelism for dynamically scaling systems in distributed environments.
%existing TPC benchmarks do not evaluate the performance of systems in managing real-time data streams. Stream processing requires benchmarks that can measure how systems process and respond to continuous input data with minimal latency.
%Moreover, (2) current benchmarks lack specific mechanisms to rigorously evaluate the efficiency of parallel and distributed processing techniques critical in stream processing architectures. Essential factors such as data streaming partitioning and required parallelism for dynamically scaling system in distributed environments are not adequately tested.}
%Stream processing systems often need to dynamically scale resources up or down based on fluctuating data volumes and computational demands. A benchmark that can simulate such environments and measure how well a system adapts to these changes is needed. 

\textbf{Machine Learning in \acrshort{dsp} systems.} In the context of benchmarking of learned component of~\acrshort{sps} such as performance prediction, existing benchmarks like DeepBench~\cite{belloni2022deepbench}, MLPerf~\cite{reddi2020mlperf}, Fathom~\cite{adolf2016fathom}, CleanML~\cite{li2021cleanml} and \added{TPCx-AI~\cite{TPCxAIBenchmark}} have laid significant groundwork. 
These benchmarks are predominantly tailored to assess specific aspects of ML systems, from the underlying hardware’s computational abilities to the effectiveness of algorithms on static dataset.
\added{For instance, MLPerf focuses on the computationally intensive aspects of ML such as model training and inference, TPCx-AI provides a more comprehensive evaluation by including the entire data processing pipeline. 
These benchmarks are valuable for organizations to optimize their end-to-end ML workflows, e.g., Dell Technologies show significant performance improvements using TPCx-AI~\cite{Nicholas2023Dell} in providing high-performance ML solutions. Their use of the benchmark has highlighted improvements in hardware efficiency and cost-performance ratios, underscoring the benchmark's utility in real-world applications.}
While existing ML benchmarks~\cite{ahmed2022runtime,boden2019benchmarking} are valuable in evaluating the capabilities and performance of various systems, a significant gap exists in terms of incorporating these findings into predictive performance models. % for future performance.} 
%However, when considering their capacity to incorporate performance prediction mechanisms directly into benchmarking processes, several nuances and limitations emerge. 
For instance, DeepBench and MLPerf primarily focus on predefined tasks for benchmarking the raw performance of hardware and machine learning frameworks %through predefined tasks, 
%They measure how well systems perform given tasks 
but do not extend to predicting future system performance based on evolving workloads or system changes. %Their static benchmarking nature provides a snapshot of performance but lacks a mechanism to use these data points for future predictions. 
%Moreover, none of the mentioned benchmarks inherently provide models to predict the future performance of systems based on historical benchmarking data. While they offer comprehensive data on various performance aspects, there is no built-in functionality to leverage this data for performance forecasting. 
Similarly, other ML benchmarks typically evaluate performance under static conditions. %and do not simulate the dynamic changes that real-world systems frequently encounter. %They fail to provide insights into how systems might perform under fluctuating workloads or infrastructure changes.
%Most ML benchmarks do not incorporate real-time data streams into their evaluations, which is crucial for systems that rely on continuous data for decision-making. Predictive analytics in such environments would benefit from benchmarks that can integrate and analyze streaming data to forecast system behavior under continuous operation.}
They do not provide insights into system performance under fluctuating workloads or infrastructure changes, nor do they incorporate real-time data streams crucial for continuous decision-making and predictive analytics in dynamic environments.

% Conclusionnuanced
\glsresetall
%\todo{add key lessons learned section and related work in the end}
\vspace{-2.5ex}
% ============================================
\section{Conclusion}
% ============================================
\vspace{-2.5ex}
This paper introduces \system that addresses the critical need for an advanced benchmarking system that reflects the complexities of modern parallel SP environments. 
Its ability to support heterogeneous hardware, integrate ML models, and evaluate a wide range of parallel query structures makes it a valuable tool for researchers and practitioners in the field of stream processing. 
%We introduced \system, a novel benchmarking system aimed at a systematic understanding of \acrshort{sps} within heterogeneous environments. % to process massive workloads from both synthetic and real-world applications. 
%Our system stands out by focusing on the complexities and dynamics of modern Stream Processing Systems
%particularly under the conditions of massive workloads that represent real-world applications. 
%Through \system, we conducted a comprehensive analysis of \textit{Flink} \acrshort{sps} and explored its performance across diverse workloads, parallel query structures and hardware configurations. 
Our findings underline the constructive impact of parallelism and hardware diversity in improving performance, but also the necessity of a refined understanding of massive parallel dataflows. % in~\acrshort{pdsp} environment for optimized performance.
In future, we plan to expand on heterogeneous hardware architectures like GPUs, followed by a in-depth analysis and training of ML models based on data collected using \system.% including the generation of diverse event rates and the simulation of real-world streaming scenarios. 

\vspace{-1ex}
\section*{\small{Acknowledgements}}
\vspace{-2.0ex}
\footnotesize{This research is funded by the DFG as part of project C2 within the Collaborative Research Center (CRC) 1053– MAKI; BMBF and the state of Hesse as part of the NHR Program and HMWK cluster project 3AI (The Third Wave of AI). We also want to thank hessian.AI at TU Darmstadt and DFKI Darmstadt.}
\vspace{-2ex}
\bibliographystyle{splncs04}
\bibliography{main}
\end{document}